\RequirePackage[hyphens]{url}
\documentclass[%
 reprint,
nofootinbib,
 amsmath,amssymb,
 aps,
english,prper,showpacs,titlepage,longbibliography,superscriptaddress,floatfix
]{revtex4-2} 

\usepackage[caption=false]{subfig}
\usepackage{graphicx}
\usepackage{dcolumn}
\usepackage{bm}
\PassOptionsToPackage{hyphens}{url}\usepackage{hyperref}
\usepackage{tabularx}
\usepackage{float}
\usepackage[dvipsnames]{xcolor}
\usepackage[T1]{fontenc}	
\usepackage[latin9,utf8]{inputenc}	
\usepackage{geometry}

\usepackage{array}
\newcommand{\PreserveBackslash}[1]{\let\temp=\\#1\let\\=\temp}
\newcolumntype{C}[1]{>{\PreserveBackslash\centering}p{#1}}
\newcolumntype{R}[1]{>{\PreserveBackslash\raggedleft}p{#1}}
\newcolumntype{L}[1]{>{\PreserveBackslash\raggedright}p{#1}}
\usepackage{comment}
\hypersetup{colorlinks=true,urlcolor=blue,citecolor=blue,linkcolor=blue}   
\usepackage[above,below]{placeins}	
\usepackage{times}
\usepackage{multirow}
\usepackage{ifthen}
\usepackage{longtable}

\usepackage{outlines}
\usepackage{adjustbox}
\usepackage{booktabs}

\newboolean{redactswitch}
\setboolean{redactswitch}{false} 

\newcommand{\redact}[1]{\ifthenelse{
    \boolean{redactswitch}}{{[Redacted]}}{
    {#1}}}

\usepackage{makecell}

\newif\ifshowtimestamp
\showtimestampfalse

\usepackage{footmisc}

\newif\ifinappendix
\let\oldappendix\appendix
\renewcommand{\appendix}{
  \oldappendix
  \inappendixtrue
}

\begin{document}

\title{Measuring student understanding in quantum computing: Development and validation of the Quantum Computing Conceptual Survey}

\author{Josephine C. Meyer}
 \affiliation{Department of Physics, California State University San Marcos, San Marcos, CA 92078, USA}
 \affiliation{Department of Physics and Astronomy, George Mason University, Fairfax, VA 22030, USA}

\author{Molly Griston}
 \email{molly.griston@colorado.edu}
\affiliation{Department of Physics, University of Colorado Boulder, Boulder, CO 80309, USA}%
 
\author{Gina Passante}%
\affiliation{%
 Department of Physics, California State University Fullerton, Fullerton, CA 92831, USA
}%

\author{Steven J. Pollock}

\affiliation{Department of Physics, University of Colorado Boulder, Boulder, CO 80309, USA}%

\author{Bethany R. Wilcox}

\affiliation{Department of Physics, University of Colorado Boulder, Boulder, CO 80309, USA}%

\date{\today}

\begin{abstract}
Research-based assessments (RBAs) have proven to be valuable tools in PER, supporting both instructional reform and foundational research. In the rapidly-growing field of Quantum Information Science (QIS), the lack of suitable RBAs limits the field's ability to make evidence-based decisions about curricula and program development. In this paper, we introduce the Quantum Computing Conceptual Survey (QCCS), an instrument developed to help address this gap by measuring student conceptual understanding in the foundations of quantum computing. Using pilot data from over 50 courses and 700 students, we present evidence supporting the validity of the QCCS for use in introductory QIS courses, drawing on analyses from both classical test theory and the Rasch model. We detail the potential uses for the QCCS, intending for it to serve as both a practical tool for instructors and a means of facilitating measurement-driven QIS education research. 

\end{abstract}

\maketitle

\section{Introduction}

Driven by the so-called Second Quantum Revolution \cite{Dowling:2003} and substantial investments such as the US National Quantum Initiative Act \cite{Raymer:2019}, we have seen the rapid development of new quantum information science (QIS) and quantum computing courses and degree programs \cite{Plunkett:2020,Aiello:2021,Cervantes:2021,Asfaw:2022,Dzurak:2022}. These courses and programs cross disciplinary boundaries and are offered at a range of degree levels \cite{Meyer:2022PhysRev}. 

The growth of dedicated QIS courses and degree programs has been accompanied by calls (e.g.~\cite{Marrongelle:2020}) to the discipline-based education research (DBER) community to intervene early in interdisciplinary QIS education while the field remains young and consensus on educational practices and curricula has yet to fully ossify. Of particular interest, QIS scientists, education researchers, and post-secondary instructors have indicated need for, and interest in, the development of research-based assessments (RBAs) targeting QIS concepts \cite{Aiello:2021, Meyer:2022PhysRev, Pina:2025}.

RBAs have a long history within the physics education research (PER) community, as well as the DBER community at large, and have proven invaluable in promoting instructional reform and enabling pedagogical research \cite{Engelhardt:2009,Hake:2011,Madsen:2017}. Benefits of RBAs range from helping instructors improve their teaching methods via reliable comparisons across instructors and institutions to validating research-based curricular materials and teaching methods \cite{Madsen:2017,Wilcox:2015Review}.

A number of RBAs have already been developed targeting various facets of student conceptual understanding in quantum mechanics courses \cite{Cataloglu:2002,Goldhaber:2009,Wuttiprom:2009,McKagan:2010,Singh:2010,Sadaghiani:2014,Marshman:2019}.
However, the content coverage and assessment objectives of these instruments are designed for a traditional quantum mechanics class offered in a physics department. Recent work has shown that the content coverage and notation in emerging interdisciplinary QIS courses differs markedly from these traditional quantum mechanics courses \cite{Meyer:2022PhysRev,Meyer:2024EPJ, Pina:2025}, rendering existing quantum assessments largely inapplicable for today's QIS courses.

In this paper, we discuss the development of the Quantum Computing Conceptual Survey (QCCS), an RBA focused on the QIS subtopic of quantum computing, the element of QIS most likely to be discussed in introductory QIS courses \cite{Meyer:2022PhysRev,Meyer:2024EPJ}. Quantum computing is generally considered the easiest QIS application to teach without a detailed background in quantum mechanics, and as such has been a traditional entry point to studying other quantum technologies. Moreover, fundamental concepts from quantum computing—such as qubits, measurement, and quantum circuits—easily transfer to other quantum information technologies.

The QCCS has been designed to target students in introductory QIS courses at the undergraduate or graduate level, given this is where many students are first exposed to QIS concepts in mathematical rigor. To our knowledge, this is the first such published instrument whose development is supported by extensive student data.\footnote{This is not to diminish the prior and ongoing efforts in QIS assessment development \cite{Durkin:2026, McGinness:2024}.} This work is also complemented by efforts including those to develop quantum computing RBAs targeted toward high school students \cite{Faletic:2023}. 

While the need for and benefit of a QIS RBA are reasonably apparent, QIS as a field of education does present several particular challenges for assessment development. First, lack of apparent consensus on the key learning goals and content of QIS education necessitated a detailed domain analysis and imposes certain content limitations, given the static nature of the instrument. Second, the inherently interdisciplinary nature of QIS coursework represents a new frontier for upper-division PER assessment given the need to construct an instrument that can be used for students with a wide range of academic backgrounds, rather than just physics students. Finally, considering the rapidly evolving nature of QIS education, it is especially important to consider how certain choices in assessment development have the potential to impact the developing field. 

This paper details the development of the QCCS and presents an initial validity argument supporting its use in QIS education. In particular, we advocate for the use of the Rasch model of measurement and adopt it as the framework of our analysis. In Section~\ref{sec:background}, we review prior work on student difficulties in quantum computing topics, discuss the history and use of measurement models in PER, and explain our rationale for using the Rasch model of measurement. Section~\ref{sec:methods} then details our methods, including initial development of the instrument, pilot testing and item refinement, and the structure of our statistical analysis. In Section~\ref{sec:analysis}, we present the results of our analysis, and Section~\ref{sec:discussion} discusses the implications for both researchers and instructors. Throughout this paper, we highlight both the strengths and novel features of this instrument, as well as the limitations of the instrument and our analysis, suggesting directions for future consideration and improvement. 

\section{Background}\label{sec:background}

In this section, we first briefly review existing work on student understanding of quantum computing fundamentals, which both relates to and helps delineate the present work. Next, we discuss the history of measurement models in PER and introduce our measurement approach, focusing on the theoretical approaches that have guided instrument development. Then, we discuss perspectives on psychometric validity to frame the presentation of our analysis. Finally, we mention potential theoretical limitations as they relate to our measurement framework. 

\subsection{Student understanding of quantum computing fundamentals}\label{sec:background_difficulties}

Research on student understanding in quantum computing has grown rapidly over the past decade, reflecting the increased adoption of quantum information science in undergraduate curricula \cite{Meyer:2026AJP}. Despite this growth, research on student understanding in quantum computing remains relatively limited compared to other topics in physics. Existing studies have begun to identify important student reasoning patterns and conceptual challenges, but they do not yet provide a comprehensive picture of learning and conceptual understanding in quantum computing. 

Much of the existing work draws on the broader literature on student understanding in quantum mechanics, as many concepts central to quantum computing—including mathematical foundations, entanglement, superposition, and measurement—have long histories of investigation within the PER literature (e.g., \cite{Marshman:2015, Kohnle:2015, Passante:2015}). At the same time, quantum computing education has developed as a distinct area of inquiry, with studies examining concepts unique or especially central to quantum computing, such as quantum gates, quantum circuit diagrams, and the distinction between classical and quantum computation (e.g., \cite{Meyer:2021PERC, Kushimo:2023, Hu:2024PRPER, Plueger:2026, Meyer:2026, Meyer:2026AJP}).

Conversely, the quantum mechanics education literature also includes many instructional contexts and concepts that are not typically emphasized in introductory quantum computing courses. For example, studies have investigated student reasoning about canonical potentials (e.g., infinite square well, harmonic oscillator, etc.) \cite{Emigh:2015}, the hydrogen atom \cite{Keebaugh:2019}, perturbation theory \cite{Kaur:2018}, and other topics that are fundamental to traditional quantum mechanics but are not central to introductory quantum computing.

The present work draws on both quantum mechanics and quantum computing education research, focusing on findings relevant to conceptual understanding in introductory quantum computing. Rather than reviewing this literature independently of the instrument, we discuss the relevant studies in the context of our construct definition in Sec.~\ref{sec:const}, specifically highlighting how prior work relates to our operationalization of the construct and the content assessed by the QCCS. 

\subsection{Measurement models in PER}\label{sec:background_measurement}

In the early days of RBA development in PER, researchers provided validity arguments through the framework of classical test theory (CTT) \cite{Engelhardt:2009}. More recently, we have seen an increased use of item response theory (IRT) frameworks, albeit often for the purpose of retrospective analysis, rather than to guide instrument development. Both CTT and IRT can be considered data-centric approaches in that validation of the instrument primarily involves fitting models to data to obtain empirically-determined item parameters. The Rasch model is mathematically equivalent to the simplest IRT model, but it differs in perspective, adopting a principles-first approach to measurement.

The Rasch model has been gaining increased traction within PER over the past several years \cite{Planinic:2019,Ding:2023}. In the context of RBAs, the Rasch model has been applied to reanalyze existing CTT-validated assessments in physics and astronomy (e.g.~\cite{Planinic:2006CSEM, Planinic:2010, Wallace:2010, Ding:2014, Susac:2018, Rainey:Thesis}). Rasch analysis has also been used to develop and validate conceptual instruments at the secondary school level (e.g.~\cite{Neumann:2013, Testa:2015, Hofer:2017, MatejakCvenic:2022, Micoloi:2025}) and recently to develop a flexible assessment of optics concepts for introductory physics \cite{Mesic:2019, SalibasicGlamocic:2021}. At the upper-division level, the Relativity Concept Inventory (RCI) \cite{Aslanides:2013}, Physics of Semiconductors Concept Inventory (PSCI) \cite{Ene:2018}, and Quantum Mechanics Evaluation (QME) \cite{ScottiDiUccio:2018} were all analyzed using the Rasch model. It is important to note, however, that applications of Rasch measurement have not been free of criticism. In particular, Ding cautions that PER literature purporting to establish Rasch measurement has sometimes failed to do so in practice, citing misapplications of theory and confirmation bias in several of the cited studies \cite{Ding:2023}.

In their development of a wave optics conceptual assessment, Mešić et al. demonstrate a practical value of the Rasch model—the ability to develop and manage item banks \cite{Mesic:2019}. Here, we would also like to emphasize the theoretical appeal to the Rasch model and the principle of specific objectivity.

\subsection{The Rasch model and specific objectivity}\label{sec:background_rasch}

Unlike data-driven psychometric approaches, the Rasch model takes an axiomatic approach and is based on an insistence on specific objectivity. Rooted in the science of metrology, specific objectivity refers to the idea that, within a frame of reference, a comparison between two students should be independent of the specific items used, and a comparison between two items should be independent of the specific students who responded to them. 

In instrument development, the goal is not to fit a generic item-response model to student data (as in IRT) but rather to construct a \textit{measurement instrument} that produces results consistent with the mathematical requirements of specific objectivity. In other words, misfit to the requirements of objective measurement demands revision of the test, not revision of the model. 

Insistence on specific objectivity demands an item-response function of the form:

    \begin{equation}\label{eq:rasch}
        P(X_{n,i} = 1)=\frac{e^{\theta_n-\beta_i}}{1+e^{\theta_n-\beta_i}},
    \end{equation}
where $P(X_{n,i} = 1)$ is the probability that student $n$, with ability $\theta_n$, correctly answers item $i$, with difficulty $\beta_i$. In this formulation, student ability $\theta$ and item difficulty $\beta$ are measured in logits (log-odds). We can see this by reformulating Eq.~\ref{eq:rasch} as:

\begin{equation}\label{eq:logit}
    \text{logit}(X_{n,i}) = \log \left({\frac{P(X_{n,i} = 1)}{P(X_{n,i} = 0)}}\right)= \theta_n - \beta_i.
\end{equation}

\begin{figure}
    \centering
    \includegraphics[width=\linewidth]{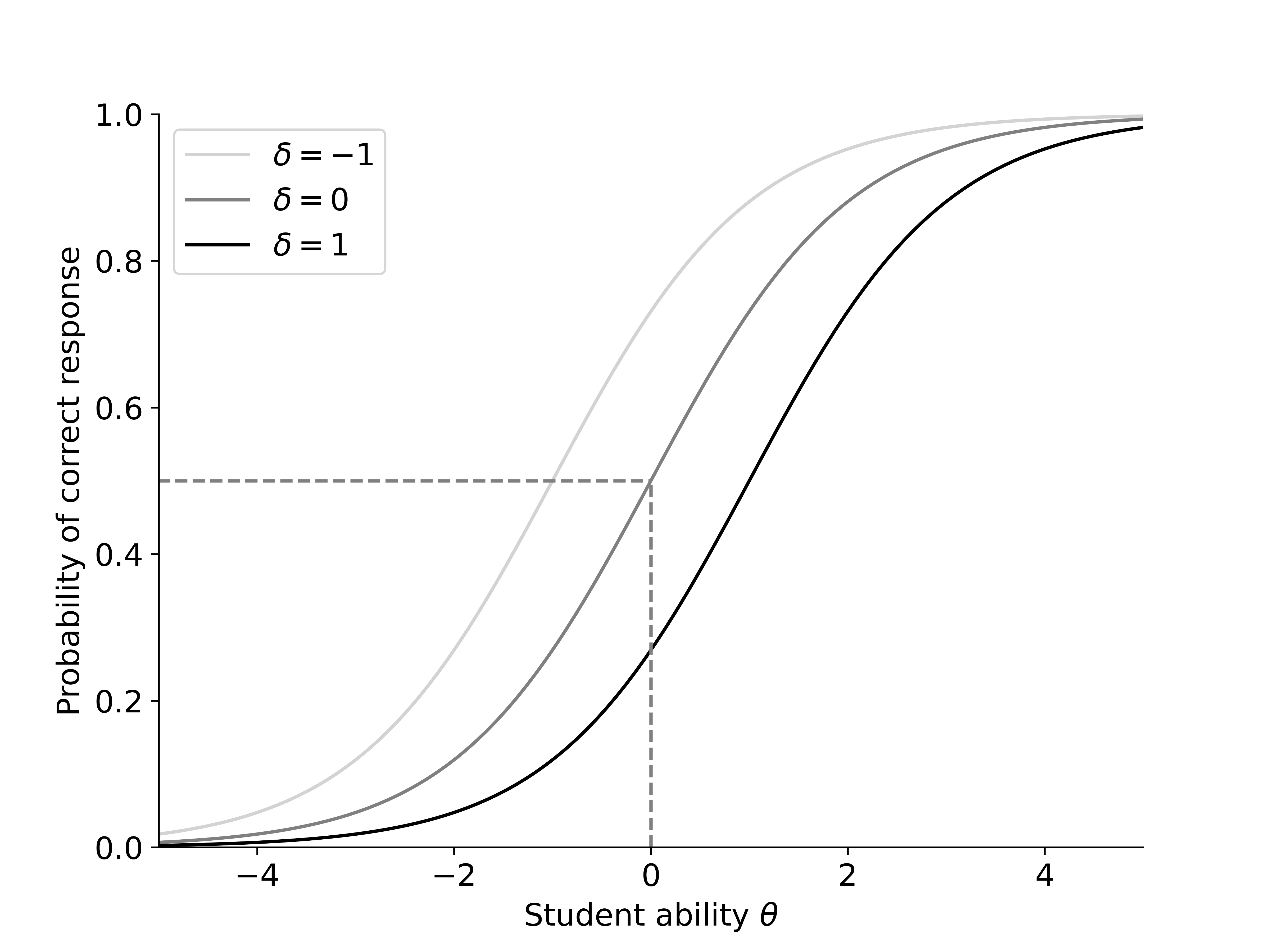}
    \caption{Example item response functions for three different item difficulties: -1, 0, 1. The dashed line indicates that a student with ability 0 will have a 50\% chance of giving the correct response on an item with difficulty 0.}
    \label{fig:ICC}
\end{figure}

This item-response function (Eq.~\ref{eq:rasch}), known as the Rasch model, is mathematically equivalent to the 1PL IRT model but derived from metrological first principles.\footnote{To further clarify, the Rasch model and 1PL IRT model are mathematically identical. Where the Rasch model differs from IRT models is in the philosophical approach to measurement: the IRT framework treats models as statistical tools, where the model is chosen to fit the data, while the Rasch measurement paradigm argues that the data must fit the model that has been derived from principles of invariant measurement.} To better understand Eqs.~\ref{eq:rasch} and \ref{eq:logit}, we consider Fig.~\ref{fig:ICC}, which shows a theoretical item characteristic curve (ICC) for three different item difficulties. By definition, a student whose ability is equal to the item difficulty has a 50\% chance of responding to the item correctly. Further, the lines all have the same slope at the inflection point, differentiating them from the ICCs of other commonly used IRT models like the 2PL and 3PL models.

The consequence of building the model based on desired properties of measurement is that an instrument exhibiting strong fit to the Rasch model has certain advantageous traits. In particular, person and item parameter estimates are mathematically decoupled, thereby producing student ability estimates that remain robust to population-wide shifts in ability distribution. Further, we can locate the person and item parameters on an invariant\footnote{It is important to note that the distance between measures is invariant (i.e., their relative values are invariant), rather than the measures themselves.} interval scale. In Section~\ref{sec:analysis}, we will provide further elaboration on how we evaluate model fit and how these parameters are estimated.

\subsection{Perspectives on validity}\label{sec:background_validity}

When developing a measurement instrument and advocating for its use, it is essential to address the instrument's validity, that is, ``the degree to which evidence and theory support the interpretations of test scores for proposed uses of the test" \cite{AERA:2014}. Unlike historical perspectives on validity that considered multiple types of validity (content, criterion-related, and construct), the modern perspective conceptualizes validity as a unified concept, where different kinds of evidence can be provided to support a validity argument \cite{Bandalos:2018}. As the above definition suggests, validity is not dichotomous, and providing evidence of validity is an ongoing process. Further, it is widely argued that it is not an instrument itself that is validated, rather it is the context, use, and interpretation of an instrument. 

The \emph{Standards for Educational and Psychological Testing} \cite{AERA:2014} describe five types of validity evidence: (1) evidence based on test content, (2) evidence based on response processes, (3) evidence based on internal structure, (4) evidence based on relations to other variables, and (5) evidence based on consequences of testing.\footnote{The inclusion of consequences of testing as validity evidence is debated; some scholars argue that consequences should inform test and score use but are not considered validity evidence.} In this paper, we will focus on the first four types of evidence, detailed in Table~\ref{tab:evidence}, and we will use these categories to help frame our discussion. However, we emphasize that these should not be viewed as supporting different types of validity, and they all contribute to our understanding of the instrument and its applications. 

\subsection{Theoretical limitations}\label{sec:background_limitations}

When developing an instrument using the Rasch model, it is important to begin with a theoretical understanding of the construct of interest. It is often suggested that the development process should begin with a definition of the construct and the creation of a concept map: an identification of levels of the construct and their corresponding behaviors \cite{Planinic:2019}. The construct map can then be used to structure item development, and once the instrument has been administered, it is possible to compare the empirical item difficulties with the predicted difficulties based on the construct map. These aspects of the development process help contribute evidence based on both test content and response processes. 

In the case of quantum computing, however, the theoretical foundations necessary to support a detailed construct map remain underdeveloped. Compared to other content areas, there is limited consensus regarding the organization of conceptual understanding and related learning progressions. Consequently, our approach to construct definition relied on triangulation across several complementary sources of evidence: a systematic domain analysis of content coverage in introductory QIS courses \cite{Meyer:2024EPJ}, instructor reports regarding important and difficult concepts \cite{Meyer:2022PhysRev}, and existing research on student reasoning and difficulties in quantum computing and related quantum mechanics contexts (e.g.,\cite{Meyer:2022PERC, Meyer:2021PERC, Kushimo:2023,Passante:2015}). Together, these sources informed our operationalization of conceptual understanding in quantum computing fundamentals used to develop the QCCS. 

As a result, we do not claim to have established a comprehensive construct map or learning progression for quantum computing. Rather, the QCCS should be viewed as an initial operationalization of the construct that is based on available empirical and theoretical evidence. Likewise, given the absence of established instruments measuring similar constructs, we do not yet report on the relationship of QCCS scores and other measures we would expect to be correlated, which is often an important source of validity evidence (evidence based on relations to other variables). As the quantum education literature continues to grow, we anticipate that more complete models of conceptual understanding, learning progressions, and complementary assessment instruments will emerge, enabling both stronger construct definitions and additional validity evidence. 

\section{Methods}\label{sec:methods}

\begin{figure}
    \centering
    \includegraphics[width=\linewidth]{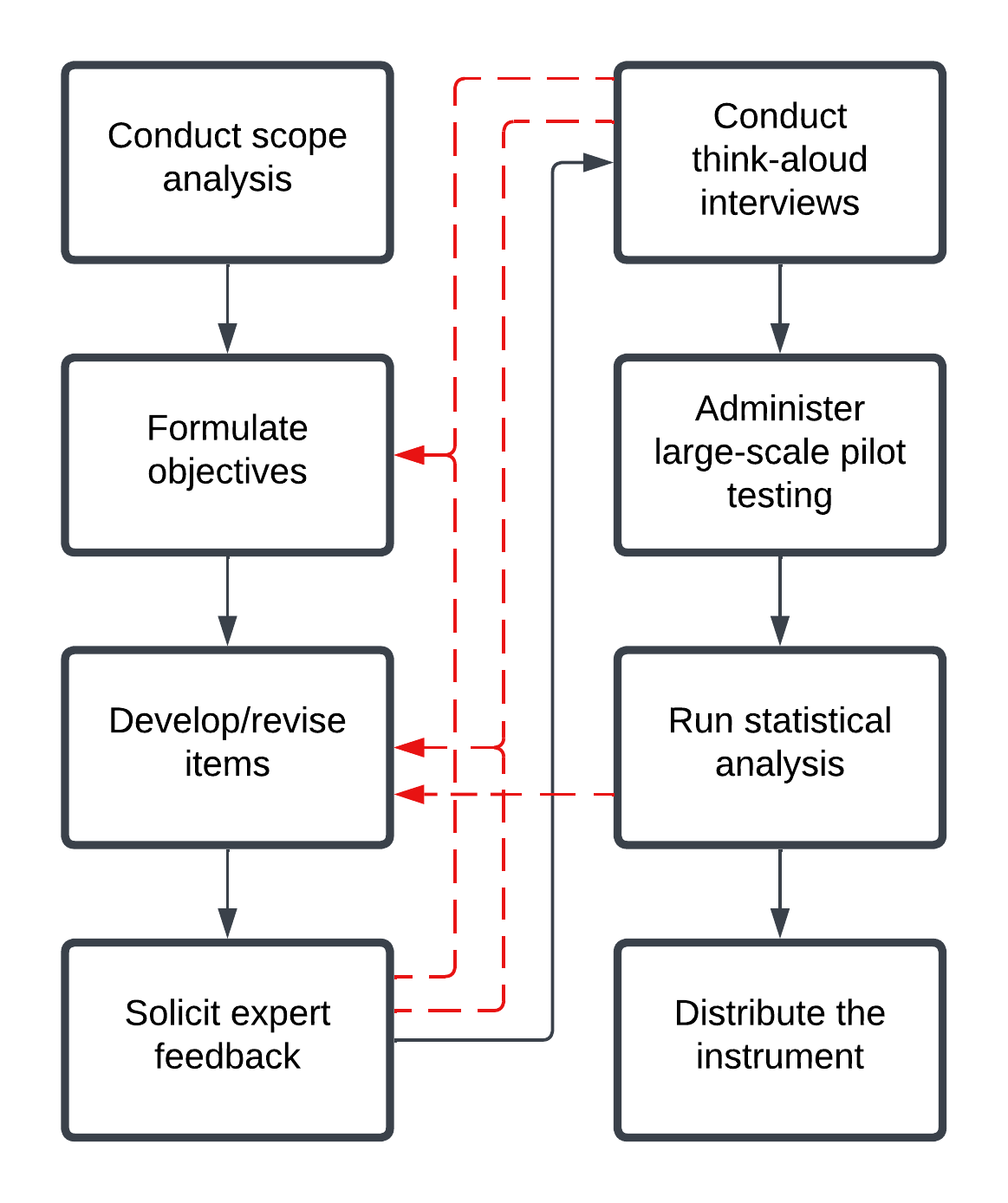}
    \caption{The general process of assessment development, where the dashed red lines indicate iteration between steps. Certain steps occurred concurrently, rather than consecutively. Figure adapted from Ref.~ \cite{Engelhardt:2009}.}
    \label{fig:process_flow}
\end{figure}

In this section, we detail the process of developing and analyzing the QCCS, following the steps illustrated in Fig.~\ref{fig:process_flow}. While the figure shows a series of consecutive steps, we note that some of these steps occurred concurrently, which will be noted where relevant. Based on the framework discussed in Section~\ref{sec:background_validity}, Table~\ref{tab:evidence} details the validity evidence presented throughout Sections~\ref{sec:methods} and \ref{sec:analysis}.

\newlength{\colA}\setlength{\colA}{0.19\textwidth}
\newlength{\colB}\setlength{\colB}{0.22\textwidth}
\newlength{\colC}\setlength{\colC}{0.45\textwidth}

\begin{table*}[tb]
\centering
\begin{tabular}{L{\colA} L{\colB} L{\colC}}
\hline\hline\\[-1ex]
\textbf{Evidence based on\ldots} &
\textbf{Validity argument} &
\textbf{Evidence obtained/presented} \\[0.5ex]
\hline\\[-0.5ex]
Test content &
Test contains a set of items that are appropriate for measuring the construct &
\textbullet\ Instructor surveys to inform topical coverage\newline
\textbullet\ Expert reviews of assessment objectives and items\newline
\textbullet\ Items developed based on research into student reasoning
\\[1ex]
Response processes &
Test items tap into the intended cognitive processes &
\textbullet\ Distractors developed from free-response items\newline
\textbullet\ Think-aloud interviews conducted with students
\\[1ex]
Internal structure &
Relations among test items mirror those expected from theory &
\textbullet\ Unidimensionality (CFA / PCAR)\newline
\textbullet\ Reliability (Cronbach's $\alpha$, McDonald's $\omega$)\newline
\textbullet\ Item--total correlations (CTT discrimination)\newline
\textbullet\ Adequate fit to Rasch model (fit statistics)\newline
\textbullet\ DIF analysis
\\[1ex]
Relation to other variables &
Relations of test scores to other variables mirror those expected from theory &
\textbullet\ Comparison of undergraduate and graduate students
\\[0.5ex]
\hline
\end{tabular}
\caption{Summary of validity evidence provided throughout the paper. This table follows Ref.~\cite{Bandalos:2018}, with the validity arguments quoted verbatim.}
\label{tab:evidence}
\end{table*}

\subsection{Construct and item development}\label{sec:methods_dev}

As discussed in Section~\ref{sec:background_limitations}, our construct definition is informed by triangulation across curriculum analyses, instructor perspectives, and existing research on student reasoning and difficulties. The remainder of this section describes how this construct was developed and operationalized in the QCCS. We first identify the content domain through a curriculum analysis, then translate this domain into measurable objectives informed by prior research on student reasoning and difficulties, and finally develop and refine assessment items aligned with those objectives.

\subsubsection{Scope analysis}

As a first step toward development of the QCCS, we conducted an exploratory survey of US QIS instructors in spring-summer 2021, receiving 27 usable responses. We conducted further focus interviews with 6 of the 27 faculty respondents. Findings from the survey and interview study are reported in Ref.~\cite{Meyer:2022PhysRev}. Instructor responses to the survey and interview prompts, contributed syllabi, and frequently-reported textbooks were then analyzed and, along with feedback from a QIS expert on our team, used to develop a follow-up survey asking faculty to report on their coverage of QIS topics. The follow-up survey was conducted in fall 2022 and received 63 responses; survey methodology and results are reported in Ref.~\cite{Meyer:2024EPJ}.

Topics were deemed viable for inclusion on the assessment if they were rated ``covered and assessed'' or ``reviewed (assume prior knowledge)'' by 80\% or more of surveyed introductory QIS instructors \cite{Meyer:2024EPJ}. Table~\ref{tab:topics-summary} summarizes the assessable topics and associated subtopics that emerged from this initial scope analysis.

During this process, we also reviewed course catalogs \cite{Cervantes:2021,Meyer:2024PRPER}, frequently-used textbooks reported in the 2021 survey (Refs.~\cite{Mermin:2007} and \cite{Nielsen:2000}), and existing quantum mechanics concept inventories \cite{Madsen:2017}.

\begin{table}[tb]
    \centering
    \begin{tabular}{l }
        \hline \hline
        \thead{Topic}\\
        \hline \\
        \textbf{Qubits} \\
        \textbf{Superposition} \\
        \textbf{Quantum gates} \\
        \ \ \ CNOT \\
        \ \ \ Hadamard (H) \\
        \ \ \ Identity (I) \\
        \ \ \ Pauli gates \\
        \textbf{Entanglement} \\
        \textbf{Quantum measurement} \\
        \textbf{Quantum circuit diagrams} \\
        \textbf{Quantum communications \& cryptography*} \\
        \ \ \ \textit{No subtopics met threshold} \\
        \textbf{Quantum algorithms*} \\
        \ \ \ Deutsch (or Deutsch-Jozsa) algorithm* \\
        \textbf{Math foundations of QIS} \\
        \ \ \ Dirac notation (bra-ket) \\
        \ \ \ Complex numbers \\
        \ \ \ Unitary matrices** \\
        \ \ \ Inner product \\
        \ \ \ Eigenvalues/eigenvectors** \\
        \ \ \ Vector spaces (finite dimensional)\\
        \ \ \ Tensor (Kronecker) product \\
        \ \ \ Dimension of Hilbert space \\
        \hline \hline    
    \end{tabular}
    \caption{Topics from the fall 2022 survey that were selected as ``covered and assessed'' or ``reviewed (assume prior knowledge)'' by at least 80\% of surveyed instructors \cite{Meyer:2024EPJ}. These topics were considered viable candidates for potential inclusion on the assessment provided they translated to measurable assessment objectives. Bolded topics represent top-level topics, with associated subtopics (if applicable) listed below. *\textit{Topics dropped from consideration in the assessment objective development stage.}
    **\textit{Topic not included on the final assessment because no items associated with this topic and its associated assessment objectives survived the iterative assessment item development process.}}
    \label{tab:topics-summary}
\end{table}

\subsubsection{Construct definition}\label{sec:const}

The scope analysis identified the broad content domain relevant to introductory QIS courses. From this domain, and with consideration of what could be meaningfully assessed using a conceptual multiple-choice instrument, we constrained the construct of interest to conceptual understanding of quantum computing; we define this as the ability to reason about the representations, properties, and evolution of single- and multi-qubit systems. Although we present these three facets separately for clarity, they are conceptually interdependent. Representations provide the formal language for the other two facets: reasoning about a property like entanglement, or about how a gate changes a system, requires first interpreting the relevant notation correctly. Evolution, in turn, describes the mechanisms by which properties change, linking the two facets together. We do not claim that these three categories constitute a complete theory of knowledge in the domain. Rather, they provide a useful preliminary operationalization of the construct.

Throughout this section, we relate our operationalization to existing literature in quantum mechanics and quantum computing education research. Much of this literature was published after the initial stages of instrument development; consequently, these studies should not be interpreted as the sole basis for item development. Rather, they are included to contextualize our construct in the relevant literature. Below, we discuss each facet, describing the relevant prior work and noting points of alignment between that work and the instrument.

\textbf{Representations of quantum systems.} This facet refers to students' ability to use, interpret, and connect the mathematical formalisms of quantum computing and connect multiple representations of quantum states and processes. This facet includes mathematical notations (Dirac notation), operations (inner products, tensor products), properties (dimensionality), and visualizations (quantum circuit diagrams).

\textit{Prior work.} Existing work in quantum mechanics education has emphasized that Dirac notation supports both efficient computation and sensemaking about abstract concepts \cite{Gire:2014, Wawro:2020, Merzel:2024}, but numerous studies have documented persistent student difficulties with Dirac notation, particularly regarding converting between Dirac notation, matrices, and wavefunctions (e.g., \cite{Singh:2013, Wan:2019}). Similar challenges have been documented in the context of quantum computing, with Kushimo and Thacker reporting student difficulties with Dirac notation specifically, as well as translating across notations \cite{Kushimo:2023}. 

Beyond notation, work in quantum mechanics education indicates that even advanced undergraduate students struggle with mathematical foundations including inner and outer products \cite{Hu:2023EJP}. In quantum computing, an especially foundational mathematical structure is the tensor product, which instructors have identified as challenging for their students \cite{Meyer:2024EPJ}. While research focusing on tensor products in this context remains limited, related difficulties have been identified in other areas of quantum mechanics education. In the context of addition of angular momentum, Zhu and Singh \cite{Zhu:2013} identified student difficulties with calculating the dimension of product states and appropriately handling operators in different Hilbert spaces.


\textit{Connection to the instrument.} This literature is broadly consistent with this facet's emphasis on multiple representations of states, gates, and circuits, and with the inclusion of tasks requiring translation across notations and representations. It is also consistent with the inclusion of tasks addressing the outcome of ``routine'' mathematical operations (inner products and tensor products), along with those targeting the dimension of multi-qubit states and the interpretation of tensor products in different contexts. For these tasks, response options include those that reflect reasoning patterns identified in this prior work (e.g., determining the dimension of the product space by summing the dimensions of the subspaces).

\textbf{Properties of quantum systems}. This facet refers to students' ability to reason about the essential features of quantum states, including superposition and entanglement. This facet encompasses both identifying these properties and using them to predict the behavior of quantum systems.

\textit{Prior work.} Existing work emphasizes that students may recognize superposition and entanglement as properties that distinguish quantum and classical computing, but they often struggle to reason about the physical implications of these properties \cite{Meyer:2022PERC, Hu:2024PRPER}. Passante et al.\ \cite{Passante:2015} found that students can use superposition ideas to calculate measurement probabilities, but they struggle to recognize how superposition states and mixed states are experimentally distinguishable; Wan et al.\ \cite{Wan:2019} found the same to be true for states differing by a relative phase. Studies have also shown that students often reason about entanglement primarily in terms of correlations between measurement outcomes \cite{Zwickl:2024}, often believing that all entangled states are maximally entangled and struggling to articulate the relationship between entanglement, product states, and separability \cite{Kohnle:2015}. 

\textit{Connection to the instrument.} Based on this work, the assessment includes tasks addressing both basic properties of superposition states and application of those properties. In particular, multiple tasks specifically probe understanding of the measurable effects of relative phases. Likewise, the instrument includes tasks that probe the ideas of entanglement, product states, and separability both independently and in combination; additional tasks require recognizing the generation and relevance of entanglement, even when it is not specifically prompted.

\textbf{Evolution of quantum systems}. This facet refers to students' ability to predict and explain how quantum systems evolve under operations and measurement. This facet includes the behavior of single- and multi-qubit gates, evolution through quantum circuits, and the effects of measurement. While closely related to the previous facet, this facet emphasizes reasoning about how those properties change under quantum operations.

\textit{Prior work.} Because quantum gates and circuits are typically specific to quantum computing courses (rather than quantum mechanics more broadly), research on this facet remains the most limited of the three. Our own prior work has shown that students often reason productively about the evolution of quantum systems by ``playing quantum computer,'' but they do not consistently apply productive conceptual resources across contexts \cite{Meyer:2021PERC, Plueger:2026}. We have also reported on student resources for reasoning about CNOT gates specifically, finding that students draw on both procedural and conceptual resources \cite{Plueger:2026, Meyer:2026}. While students were found to use qualitative rules about control and target qubits productively, these ideas also led to incorrect conclusions when overgeneralized \cite{Plueger:2026}. In particular, students struggled with recognizing that CNOT gates can generate entanglement \cite{Meyer:2026}. 


\textit{Connection to the instrument.} The instrument includes both procedural gate-application tasks and those requiring the application of gates in more conceptually involved contexts. 

To guide item development, we translated these facets into a set of assessment objectives (AOs) describing the specific abilities the instrument should measure \cite{Vignal:2022}. These objectives, detailed in Appendix \ref{sec:appAO}, served as a bridge between the construct definition and the development of assessment items. Many items on the instrument require an intersection of these facets, consistent with the idea that they are intended to represent interrelated parts of a single construct, rather than separate dimensions. For example, Item 16 asks students if a given gate, represented in a quantum circuit diagram (QCD), produces entanglement between two qubits. To answer this question, students must read a QCD and then coordinate ideas about the behavior of quantum gates, the interpretation of tensor products, and the characteristics of entanglement.

\subsubsection{Item development and refinement}

Initial drafts of items were developed in 2023 based on the AOs. Most items were initially written as free-response items with the intention of eventually making them closed form to facilitate scalability of the assessment. As detailed in Section~\ref{sec:const}, item development was informed by early research into student reasoning in quantum computing \cite{Meyer:2021PERC,Meyer:2022PERC,Kushimo:2023}, student difficulties reported by instructors \cite{Meyer:2022PhysRev}, and pedagogical experience by members of the research team. We also reviewed the PER literature on student difficulties and reasoning patterns in traditional quantum mechanics and identified trends from the literature that appeared likely to carry over into the quantum computing context \cite{Singh:2006,Passante:2015,Kohnle:2015,Wan:2019}. Initial item drafts were refined and the most promising items were identified through discussions among the research team. 

\subsection{Interviews and pilot testing}

Once the items were initially developed, they went through many rounds of revision based on instructor and student interviews, as well as large-scale pilot testing. Details about the rounds of pilot testing and pilot course characteristics can be found in Tables~\ref{tab:pilot-sites} and~\ref{tab:pilot-sites-demographics}. We note that our pilot courses are primarily housed at Public R1 Universities, and approximately 20\% of them are at underrepresented minority (URM)-serving institutions. The pilot courses serve both undergraduate and graduate students—frequently in the same classroom—and are primarily listed in physics, computer science, and electrical and computer engineering departments. Details about the full student population characteristics can be found in Appendix~\ref{sec:appA}.

During fall 2023, we conducted interviews with $N=6$ faculty with experience in QIS education. To ensure a range of perspectives, the faculty interviewed intentionally included both technical QIS experts (i.e.\ researchers) and non-experts who had self-studied the material to teach it. Faculty were first shown the list of draft AOs and asked to comment on the relevance, clarity, and importance of the AOs. Faculty were then asked to comment on the draft assessment items, including whether the items aligned with the relevant AOs, used content and notation that students in their courses could expect to see, and avoided notational or verbal ambiguity. 

Through instructor feedback, we heard that academic disciplines (especially physics vs.\ computer science) tended to vary in qubit endianness and indexing,\footnote{In this context, endianness refers to whether the qubit corresponding to the least significant classical bit is written on the right/left of a tensor product or top/bottom of a circuit diagram. Indexing refers to how qubits are numbered, whether starting from 0 or 1.} requiring a number of assessment items to be restructured or eliminated to avoid biasing in favor of one notational convention. Faculty input also led to minor revisions in the wording of the AOs and test items and assisted with selection of items to advance from v1.0 to v2.0 of the assessment.

In addition to interviews with instructors, we conducted think-aloud interviews with students to ensure they were interpreting the items and distractors as intended. We conducted interviews with $N=6$ students or former students at a large, public R1 university who had taken an undergraduate quantum computing course in the previous 2 academic years. The questions in these interviews were taken from QCCS v1.0, in some cases trialing modifications or adaptations intended for v2.0 based on preliminary analysis. 

In the same semester (fall 2023), we conducted the first round of large-scale pilot testing using v1.0. Pilot sites were recruited through direct email solicitation of faculty identified as teaching a QIS course at a US institution, as determined by analysis of publicly-available course schedules. Faculty were promised their course's results, as well as a comparison with other pilot sites, as incentive to participate. 

The first version of the assessment (v1.0) featured 24 primarily free-response items. To keep assessment length manageable and minimize survey fatigue, the assessment was split into 8 anchor items shown to all students, 8 items on form A, and 8 items on form B. After completing the anchor items, students were randomly assigned either form A or form B, with the option to complete the remaining questions for additional practice afterward.

Results were analyzed using classical test theory to obtain preliminary estimates of item and test parameters.\footnote{For the first round of pilot testing, exploratory Rasch modeling was also conducted; however, the combination of non-dichotomous items and high proportion of misfitting items limited interpretability at this stage.} Differential item functioning was also tested for, but the small sample size limited statistical power at early stages. 

Based on results from the first pilot administration and instructor and student interviews, we developed v2.0 of the QCCS. Version 2.0 included 20 closed-form items, with distractors chosen based on common student difficulties in the free-response versions; v2.0 was pilot tested during spring 2024, and results were comprehensively analyzed using both CTT and Rasch modeling.

With preliminary results from v2.0, we identified several potentially problematic items on the instrument and conducted interviews with $N=30$ students, focusing on these items \cite{Plueger:2026}. The interviewed students represented coverage across Physics, Engineering, Computer Science, Math, and QIS-specific majors, and spanned all undergraduate and graduate academic levels; the students were majority Asian and Male. Based on this second round of student interviews, items were refined, and in a few cases, wholly restructured (e.g., \cite{Meyer:2026}).

Version 2.1 was piloted in fall 2024. The results of v2.1 were used to make small modifications to six items (e.g., simple wording changes) and a more substantial modification to one item.

With these edits made, v2.2 was piloted in spring 2025, with data collection continued in fall 2025. In this paper, we report the results from v2.2, the current version of the instrument. 

\begin{table*}[tb]
    \centering
    \begin{tabular}{l c c c c}
    \hline \hline
         \thead{Version} & \thead{1.0} & \thead{2.0} & \thead{2.1} & \thead{2.2}\\ \hline
         \thead{Semester} & Fall 2023  & Spring 2024 & Fall 2024  & Spring/Fall 2025 \\
         \thead{Item Format} & FR \& MC & MC & MC & MC \\
         \thead{Num. total items} & 24* & 20 & 20 & 20 \\
         \thead{Num. unique courses} & 19 & 43 & 29 & 55\\
         \thead{Num. unique institutions} & 18 & 40 & 28 & 46\\
         \thead{Num. usable responses} & 271** & 621 & 345 & 777 \\
         \hline \hline
    \end{tabular}
    \caption{Summary statistics for each of the four pilot semesters of data collection. Items formats include free-response (FR) and multiple-choice/multiple-select (MC). Usable responses indicates the number of unique students whose responses were usable for statistical analysis after cleaning and filtering the data. *Includes 8 anchor items, 8 on form A, and 8 on form B randomly assigned to students. **Refers to the number of unique respondents: includes 170 usable responses to the Form A questions, and 163 usable responses to the Form B questions (these values do not sum to 271 because some students chose to respond to both forms).}
    \label{tab:pilot-sites}
\end{table*}

\begin{table*}[]
    \centering
    \begin{tabular}{l c c c c}
        \hline \hline
         \thead{} & \thead{v1.0} & \thead{v2.0} & \thead{v2.1} & \thead{v2.2}\\
         \hline
         Unique courses     & 19        & 43        & 29 & 55\\
         \multicolumn{4}{l}{\ \ \textit{Institutional characteristics}} \\
         Minority serving   &  4 (21\%) & 11 (26\%) &  6 (21\%) & 16 (29\%)  \\
         URM-serving\footnote{As in Ref.~\cite{Meyer:2024PRPER}, we define a URM-serving institution as a minority serving institution \cite{NASA:2023, Carnegie:2025} that is not solely an Asian American and Native American Pacific Islander-Serving Institution (AANAPISI).}        &  4 (21\%) &  9 (21\%) &  5 (17\%) & 12 (22\%)\\
         R1                 & 15 (79\%) & 34 (79\%) & 23 (79\%) & 46 (84\%)  \\
         R2                 &  1 (5\%)  &  5 (12\%) &  3 (10\%) & 5 (9\%)\\
         RCU\footnote{Research College and University (RCU) was first introduced in the 2025 Carnegie Classification System \cite{Carnegie:2025}.} &
         0 (0\%)  &  0 (0\%) &  0 (0\%) & 3 (5\%)\\
         Non-research       &  3 (16\%) &  4 (9\%)  &  3 (10\%) &1 (2\%)\\
         Public             & 14 (74\%) & 29 (67\%) & 20 (69\%) & 35 (64\%)\\
         \multicolumn{4}{l}{\ \ \textit{Course level}} \\
         Undergraduate      & 16 (84\%) & 28 (65\%) & 22 (76\%) & 42 (76\%)\\
         Graduate           &  7 (37\%) & 27 (63\%) & 16 (55\%) & 32 (58\%)\\
        \multicolumn{4}{l}{\ \ \textit{Course departmental listing}} \\
        Physics             & 13 (68\%) & 20 (47\%) & 19 (66\%) & 32 (58\%)\\
        Computer science    &  4 (21\%) & 17 (40\%) &  9 (31\%) & 23 (42\%)\\
        ECE                 &  5 (26\%) & 15 (35\%) &  8 (28\%) & 17 (31\%)\\
        Math                &  1 (5\%)  &  4 (9\%)  &  2 (7\%)  & 5 (9\%)\\
        Other               &  0 (0\%)  &  3 (7\%)  &  1 (3\%)  & 5 (9\%)\\
        \hline \hline
         
    \end{tabular}
    \caption[Demographic profiles of the QCCS pilot courses.]{Demographic profiles of the QCCS pilot courses. Statistics are reported out of the total number of courses (not institutions); some institutions may be counted more than once due to offering multiple courses. Percentages may not add up to 100\% due to rounding and/or non-exclusive categories.}
    \label{tab:pilot-sites-demographics}
\end{table*}

\subsection{Statistical analysis}\label{sec:methods_stat}

Here, we outline the quantitative approaches used to characterize the measurement properties of the instrument and support the validity of score interpretations; further methodological specifications will accompany the results in Section~\ref{sec:analysis}. 

Our statistical analysis focuses on evidence based on internal structure and relations to other variables, complementing the qualitative evidence based on test content and response processes discussed in Sec. \ref{sec:methods_dev}. We draw on both CTT, to align with existing work in PER, and Rasch-based methods, for the reasons discussed in Section~\ref{sec:background_measurement}.

We first evaluate the dimensionality of the instrument to determine whether responses reflect a single underlying construct that supports the interpretation of an overall score. Dimensionality is assessed using both factor-analytic and Rasch-based methods to provide complementary perspectives on the internal structure of the instrument. This step provides evidence toward validity based on internal structure and also tests assumptions of both CTT and the Rasch model.

We then provide item- and test-level characteristics using a CTT framework. Item difficulty and discrimination indices are used to determine whether items span an appropriate range of the construct and effectively distinguish between respondents with high and low levels of the underlying trait. The reliability of the responses describes the proportion of variance in observed scores that can be attributed to variance in true scores, rather than error or noise; an estimate of reliability is important in ensuring the consistency of scores and describing measurement error. In addition, we evaluate test-level discrimination to consider how effectively the instrument differentiates among respondents in the sample population. 

We next present a Rasch-based analysis to provide a model-based examination of measurement properties. We first assess the model assumptions, considering item independence in addition to unidimensionality. We then evaluate model fit to ensure the observed data conform to the Rasch model before estimating item difficulties and person abilities. These parameters are then used to create a Wright map (a graphical representation of item difficulties and person abilities) to consider whether the items provide sufficient coverage and appropriately target the intended population. Additionally, we consider the distributions of ability for students at different academic levels to assess whether group differences align with theoretical expectations. Finally, we examine test information functions, standard error of measurement, and Rasch-based reliability indices to evaluate the precision of measurement across levels of the construct. Together, these analyses provide an evaluation of whether the instrument yields internally consistent, well-targeted, and interpretable scores that can support valid inferences about respondents' quantum computing conceptual ability. 

\subsubsection{Differential item functioning}

One essential characteristic of a high-quality test is fairness of the test across social groups. Fairness is necessary to ensure research findings based on the QCCS generalize across populations, and that the instrument itself is usable by instructors with students of different backgrounds (a concern in PER given that student pilot pools have disproportionately skewed white, male, and toward relatively elite institutions \cite{Kanim:2020}). Fairness is also a validity concern: if performance on a test item depends significantly on an extraneous factor (e.g., race), the test's validity as a measurement instrument is called into question as inter-group comparisons (e.g., between the student bodies at minority-serving and predominantly-white institutions) become difficult or impossible.

Within a psychometric framework, the primary way to evaluate fairness of an instrument is through the examination of differential item functioning (DIF). DIF is defined as ``a factor other than ability... affect[ing] the likelihood that a student will answer the question correctly'' \cite{Dietz:2012}. DIF analysis compares student response patterns to individual items across identity groups while controlling for students' overall ability. Beyond flagging potential equity issues, studies of DIF contribute validity evidence based on internal structure, as they evaluate whether or not items behave in a way that aligns with existing theory.

As with all statistical analysis, DIF signals ought to be considered holistically; not all DIF is an equity or validity concern. For instance, in our context, DIF across students of different academic backgrounds (primarily major and previous coursework) could provide information about the construct of interest and is not evidence of inherent flaws in the instrument.\footnote{It is still important to identify and understand such DIF, particularly when intending to make inter-group comparisons.} 

Throughout item development and revision, we have conducted preliminary DIF analyses and prioritized revising items exhibiting problematic DIF. In our future work, we intend to conduct an extensive DIF analysis considering potential differences across academic backgrounds, as well as intersectional effects. However, due to sample size limitations, the DIF analysis in this paper will focus only on the axes of race and gender. 

\section{Analysis and Results}\label{sec:analysis}

In this section, we will first describe the process of data cleaning and then follow the statistical analysis outline described in Section~\ref{sec:methods_stat}.

\subsection{Data cleaning}

Before beginning our analysis, we removed responses where students did not consent to the sharing of their data with the research team. Next, duplicate responses by the same student were flagged and removed; the most complete response was kept, or in the case of multiple complete responses, the first complete response. Incomplete responses (defined as $\le$75\% of items viewed) were also dropped at this time, along with responses where students did not respond to at least half of the questions. Write-in answers to optional demographic questions were also recoded where applicable.

\subsubsection{Rapid-guessing identification}

When administering RBAs, it is important that students are not awarded grades for correctness \cite{Madsen:2017}. Instead, incentivization in the form of participation credit or extra credit is highly recommended to maximize participation rates. However, incentives combined with the low-stakes nature of the assessment can lead to another problem: students randomly clicking through the survey to receive participation credit. We refer to such behavior as ``rapid-guessing" behavior, and it is prevalent in many low-stakes assessment contexts \cite{Wise:2017}. The inclusion of such responses can affect student ability estimates and item parameter estimates, and it is thus important to identify rapid-guessing behavior prior to analysis. 

\begin{figure*}[t]
    \centering
    \includegraphics[width=\linewidth]{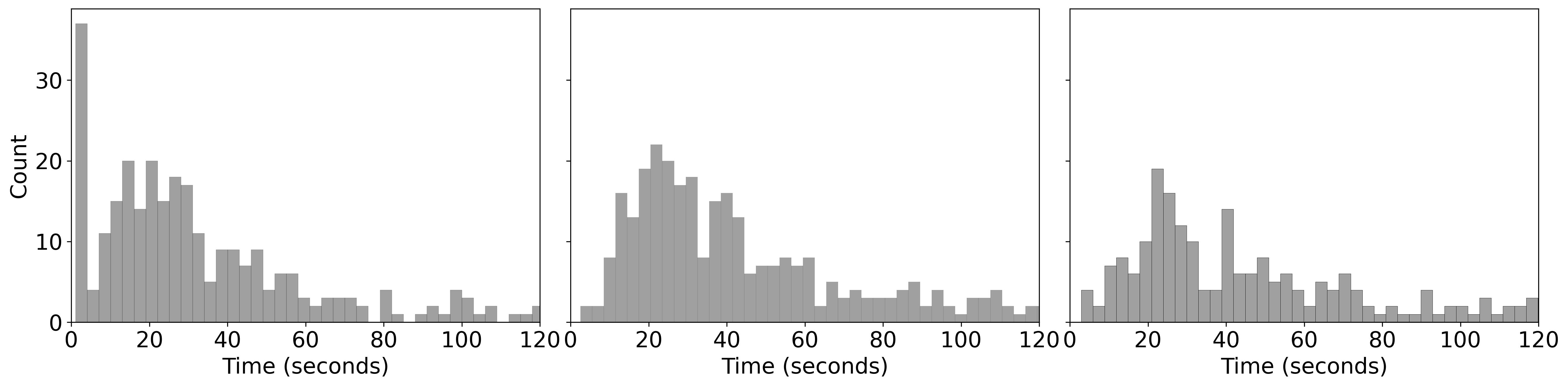}
    \caption{Response time distribution for a sample item, where students have been binned into three quantiles based on overall score (increasing left to right).}
    \label{fig:rt_thirds}
\end{figure*}

To facilitate this filtering, we collected item response times (i.e., how long was spent on each question) as students were completing the survey. Since students are able to leave the page while taking the survey, item response time is an imperfect metric; however, it can still be used to help filter out students exhibiting rapid-guessing behavior. As demonstration of this, Fig.~\ref{fig:rt_thirds} shows the response time distribution for a sample item, with the respondents split into three groups by total score. We see students in the bottom third engaging in rapid-guessing behavior, evidenced by a very short response time. Based on literature recommended practices \cite{Wise:2017}, we identified a rapid-guessing threshold for each item and then, for each student, calculated a response time effort (RTE) index—the percentage of a student's responses that do \emph{not} exhibit rapid guessing behavior. Having compared the performance of several rapid-guessing threshold identification methods, we chose to use a constant 5-second threshold for each item. For filtering students based on RTE, a commonly used threshold is 0.9 \cite{Rios:2021}, meaning that students will be filtered out if they showed evidence of rapid guessing behavior on greater than 10\% of questions. Based on our RTE distributions and considering the fact that our RT data is less reliable than the common context of large-scale assessments (where students must complete the assessment in one sitting), we choose a slightly less strict cutoff RTE of 0.8. This corresponded to the removal of 6.5\% of the sufficiently complete responses. 

\subsection{Dimensionality of QCCS}
\label{sec:unidimensionality}

To assess the dimensionality of the instrument and consider whether the requirement of unidimensionality is sufficiently met, we use both factor-analytic and Rasch-based approaches. 

\subsubsection{Single factor CFA}

First, we conduct a confirmatory factor analysis (CFA) based on our hypothesis of a single factor structure. This methodology is based on a structural equation modeling framework and is common in the PER literature (e.g., \cite{Eaton:2018, Cioffi:2024}). Essentially, the CFA evaluates the goodness of fit of our hypothesized one-factor model. Since the observed estimators are dichotomous, we use the weighted least squares mean and variance adjusted (WLSMV) estimator. To evaluate the model fit, we use scaled versions of the comparative fit index (CFI), the Tucker-Lewis fit index (TLI), standardized root mean square residual (SRMR), and root mean square error of approximation (RMSEA). With dichotomous variables, there is no consensus about the thresholds necessary to identify ``good fit'' of a model \cite{svetina2020multiple}, but we will use the following criteria as a guideline: CFI $\geq$ 0.95, TLI $\geq$ 0.95, RMSEA $\leq$ 0.05, SRMR $\leq$ 0.08 \cite{Hu:1999}. 

With a one-factor model, our scaled global fit statistics are: CFI = 0.964, TLI = 0.960, RMSEA = 0.036, and SRMR = 0.066. Based on our discussed cutoffs, these indicate good global fit. To evaluate local fit, we consider the residual matrix, and we do not observe any evidence of additional dimensions. For the single-factor structure, all of the factor loadings  can be found in Table~\ref{tab:loadings}. Though practices vary in the literature, the rule of thumb is that factor loadings of 0.3 or higher are sufficient for measurement \cite{Hair:1995}. For the one-factor model, observe that all factor loadings are above 0.3, with the lowest loading being 0.38 for Item 8.

\begin{table}[h]
    \centering
    \begin{tabular}{c c | c c}
    \hline \hline
    \thead{Item} & \thead{Loading} & \thead{Item} & \thead{Loading} \\
    \hline
    \textbf{1}  & 0.58 & \textbf{11} & 0.61 \\
    \textbf{2}  & 0.65 & \textbf{12} & 0.59 \\
    \textbf{3}  & 0.66 & \textbf{13} & 0.70 \\
    \textbf{4}  & 0.55 & \textbf{14} & 0.52 \\
    \textbf{5}  & 0.58 & \textbf{15} & 0.55 \\
    \textbf{6}  & 0.58 & \textbf{16} & 0.68 \\
    \textbf{7}  & 0.52 & \textbf{17} & 0.70 \\
    \textbf{8}  & 0.38 & \textbf{18} & 0.50 \\
    \textbf{9}  & 0.70 & \textbf{19} & 0.63 \\
    \textbf{10} & 0.76 & \textbf{20} & 0.47 \\
    \hline \hline
    \end{tabular}
    \vspace{10pt}
    \caption{Factor loadings for the one factor model.}
    \label{tab:loadings}
\end{table}

We find the results of the single-factor CFA to show evidence of sufficient unidimensionality to merit using a single sum score.

\subsubsection{PCAR}

In addition to the CFA, we conduct a principal components analysis of residuals (PCAR) after fitting the Rasch model, which is an approach more in line with the Rasch model framework. Conceptually, the PCAR assesses whether there are remaining correlations after fitting the Rasch model that may be evidence of additional dimensions. The literature recommends investigating potential additional dimensions if the eigenvalue of the first contrast (i.e., the largest eigenvalue) is greater than 2 \cite{Linacre:PCAR}. Otherwise, it is reasonable to argue that the residuals are explained by random noise, rather than additional dimensions. A PCA of the standardized residuals yields a largest eigenvalue of 1.57. As such, the PCAR provides further support of sufficient unidimensionality, and we consider it reasonable to proceed with using the Rasch model.  

\subsection{CTT Analysis}

\subsubsection{Item difficulty and discrimination}
\label{sec:results-CTT} \label{sec:ctt}

\begin{figure*}
    \centering
    \includegraphics[width=0.8\linewidth]{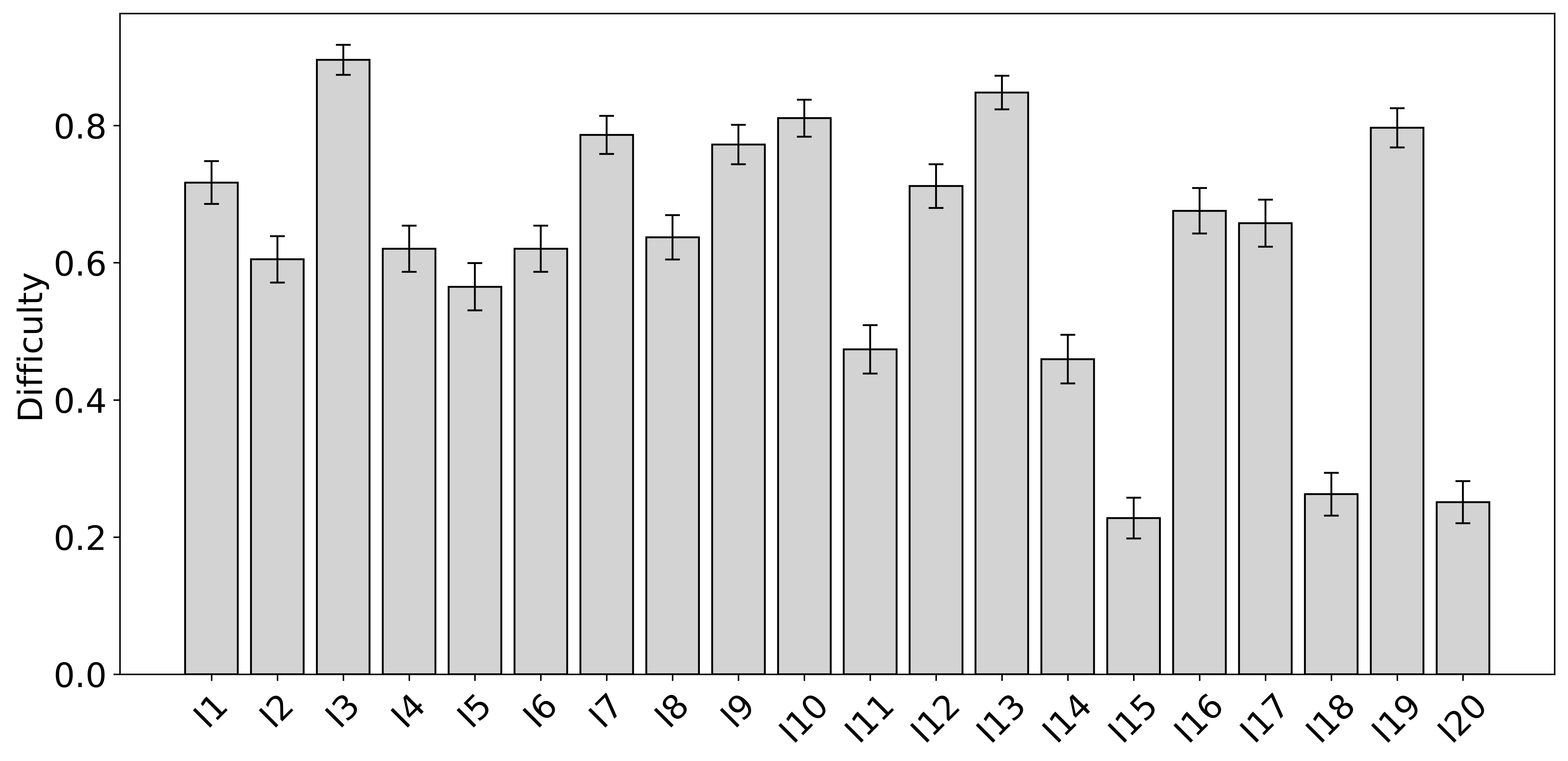}
    \caption{CTT item difficulties for v2.2. The error bars represent a 95\% confidence interval. Numerical values are given in Table~\ref{tab:ctt-diff-disc}.}
    \label{fig:ctt_difficulty}
\end{figure*}

We first present the item level characteristics, beginning with item difficulty and discrimination. An item's difficulty is simply the proportion of respondents who answered the item correctly. Previous work in PER has recommended maintaining item difficulties between 0.3 and 0.9 \cite{Engelhardt:2009}, as items that are either very easy or very difficult provide minimal information for the majority of students. However, we are also attentive to ensuring both sufficient topical coverage and the instrument's ability to discern across a wide range of student abilities. Additionally, many of the QCCS items have structurally very low guessing floors compared to traditional multiple-choice questions. As such, we are not concerned about items that are slightly harder than the originally suggested threshold. Figure \ref{fig:ctt_difficulty} shows the item difficulties, with all difficulties falling between 0.2 and 0.9.

\begin{figure*}
    \centering
    \includegraphics[width=0.8\linewidth]{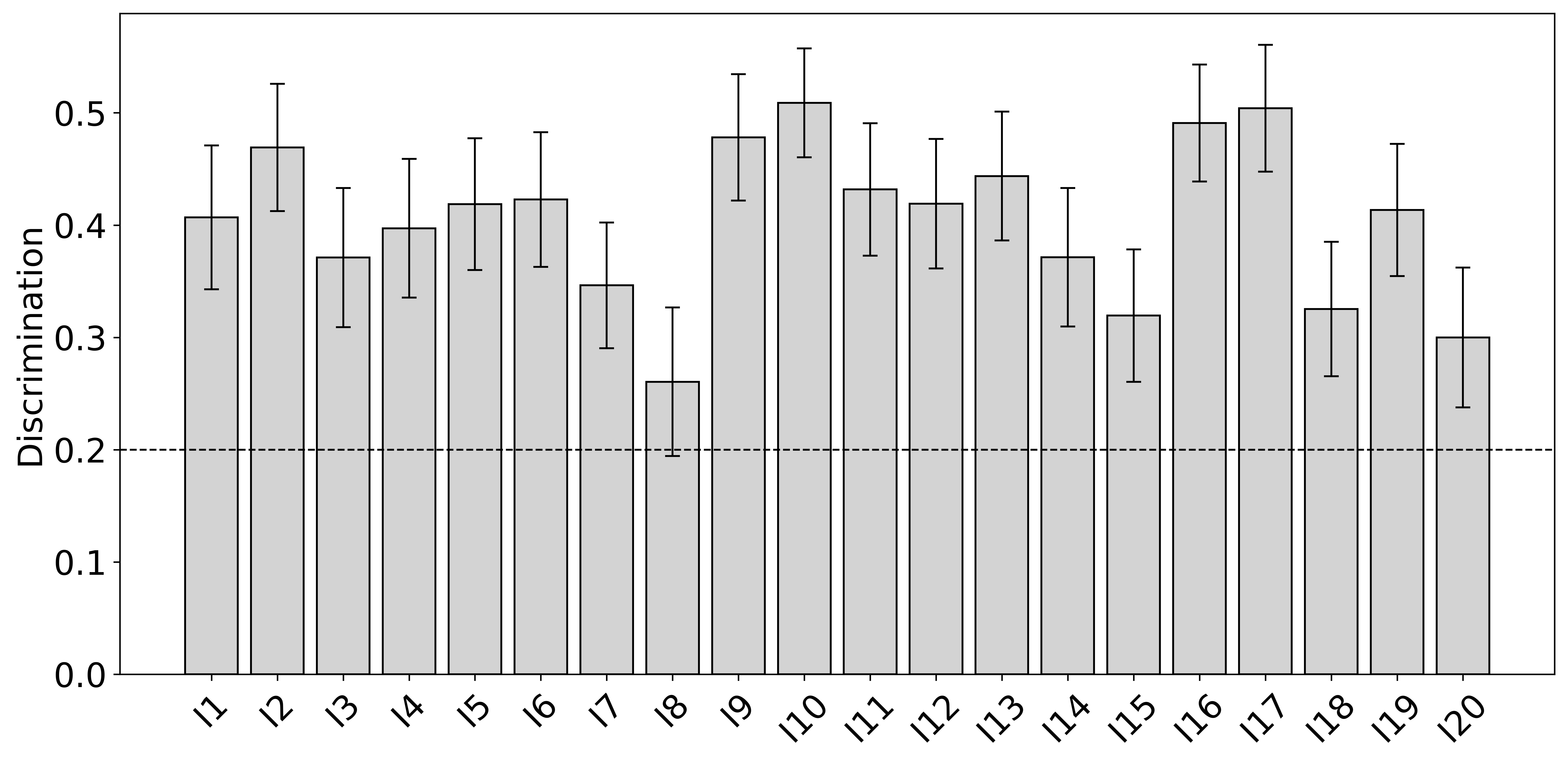}
    \caption{CTT item discriminations v2.2. Item discrimination is calculated as the point-biserial correlation, and the error bars represent a 95\% confidence interval. The dashed line represents the literature-suggested threshold of 0.2. Numerical values are given in Appendix~\ref{sec:appB}.}
    \label{fig:ctt_discrimination}
\end{figure*}

Item discrimination refers to an item's ability to discern between high and low-performing students, and it is calculated as the correlation between item score and total score. Specifically, we report the point-biserial correlation coefficient, with auto-correlation dropped, which is a special case of the Pearson correlation coefficient in which one variable is continuous and the other is dichotomous. The literature suggests that the point-biserial correlation coefficient should be at least 0.2 for all items \cite{Engelhardt:2009}. Figure \ref{fig:ctt_discrimination} shows the item discriminations, with all items having a discrimination between 0.26 and 0.51, indicating sufficient discrimination based on CTT criteria. Both CTT item difficulty and discrimination values can also be found in Appendix~\ref{sec:appB}. 

\subsubsection{Whole-test reliability and discrimination}

Regarding test-level statistics, we first focus our attention on measures of reliability. Historically, reliability has been estimated using Cronbach's $\alpha$; however, many recent studies have drawn attention to issues with Cronbach's $\alpha$ \cite{Hayes:2020}. In particular, the calculation of $\alpha$ assumes essential tau-equivalence (i.e., equal factor loadings across items), which is often not a reasonable assumption to make. In replacement, methodologists have suggested the use of McDonald's $\omega$, which is a more general form of $\alpha$ that does not assume tau-equivalence. Here, we report both $\alpha$ (for familiarity) and $\omega$. Within PER, researchers have advised a reliability of at least 0.7 for group measurements, and at least 0.8 to be potentially used for individual measurements \cite{Engelhardt:2009}. We calculate $\alpha$ and $\omega$ to both be 0.83, which is above the discussed threshold. We consider them evidence that our instrument is sufficiently reliable for our goals. 

In the past, researchers have also reported a test-level discrimination index, Ferguson's $\delta$, which aims to describe the overall discrimination. The literature has suggested that an acceptable value for Ferguson's delta is $\delta>0.9$ \cite{Engelhardt:2009}.\footnote{We note that Ferguson's $\delta$ has also been the subject of criticism, namely that it is a characteristic of the population and does not provide useful information about a measurement instrument. Again, we will report Ferguson's $\delta$, but we caution the reader when making interpretations.} Here, $\delta = 0.98$, which is above the discussed threshold and does not cause reason for concern.

\subsection{Rasch analysis}
\label{sec:results-Rasch}

Next, we present the results of our Rasch-based analysis. All calculations in this section were performed using R's \textbf{TAM} package library \cite{TAM} using marginal maximum likelihood estimation, except where explicitly declared otherwise. 

\subsubsection{Rasch model assumptions}

The Rasch model makes two important assumptions: unidimensionality and local independence. The condition of unidimensionality was verified in Sec.~\ref{sec:unidimensionality}. The assumption of local independence means that a student's responses to any two items on the instrument should not be related to one another, except through the student's ability. There are a number of ways to evaluate local independence. First, items showing evidence of ``overfit'' can be evidence of a lack of local independence—this will be discussed in more detail in the following section about fit statistics. Another method of evaluating this requirement is to calculate Yen's Q3 statistic, which considers the correlation between raw item residuals. Previous work has used a wide range of thresholds for Yen's Q3 statistic, but a commonly used procedure is to examine pairs of items with a statistic greater than 0.2 for potential violations of local independence \cite{Christensen:2017}. 

In our data, we observe two item pairs that have adjusted Q3 statistics with an absolute value greater than 0.2: Item 3 and Item 19 have an adjusted statistic of 0.24, and Item 8 and Item 15 have an adjusted statistic of -0.23. For the former pair, both items ask about the form of a tensor product, so it is understandable that we observe a higher correlation; however, one asks about the tensor product of two states, and the other asks about the tensor product of two gates. We consider these conceptually different enough to justify retaining both items. For the latter pair, considering the content of the items and the fact that the residual correlation is negative, we are not concerned that this is evidence of a violation of local independence. For all of these items, we also evaluate the item fit statistics to ensure the items are not problematic. 

\subsubsection{Item fit}

Having addressed the requirements of the Rasch model, the next task is to evaluate item fit. As discussed in Sec.~\ref{sec:background_rasch}, we must assess whether the data fit the model. To do so, we use chi-square fit statistics; specifically, we consider infit and outfit mean square statistics. Both infit and outfit are based on the sum of squared standardized residuals, but infit is information weighted, meaning more weight is given to well-targeted observations (observations for which the student ability is close to the item difficulty). Both statistics have an expected value of 1 and can range from 0 to positive infinity. Here, it is important to emphasize that the Rasch model is a probabilistic, or stochastic, model and therefore predicts a certain amount of variation. A fit statistic of 1.2 can be interpreted as the data having 20\% more variation than predicted by the Rasch model, and a fit statistic of 0.8 can be interpreted as the data having 20\% less variation than predicted by the model. The former is known as "underfit" and the latter as "overfit." 

When evaluating fit statistics, overfit can be evidence of item dependence—the responses are too predictable because the items are not actually independent. Bond et al.\ \cite{Bond:2020} argue that overfit often has no practical consequences, and the technical consequences are limited to artificially small standard errors and high reliability. Underfit, however, is considered  potentially harmful to measurement, and can be evidence of poorly written items or undesirable respondent behavior (e.g., guessing). Typically, studies using the Rasch model cite a range of ``acceptable" fit statistics; however, Ding notes a lack of consistency both between studies and within studies in PER regarding what is considered an acceptable range for identifying misfitting items \cite{Ding:2023}. Bond et al. \cite{Bond:2020} advise a nuanced interpretation of fit statistics but consider [0.8, 1.2] to be a reasonable range for high stakes multiple-choice tests and [0.7, 1.3] for low stakes multiple-choice tests. In our design, we have targeted a range of [0.7, 1.3], but we will discuss items falling near the edge of that range. 

\begin{figure}
    \centering
    \includegraphics[width=\linewidth]{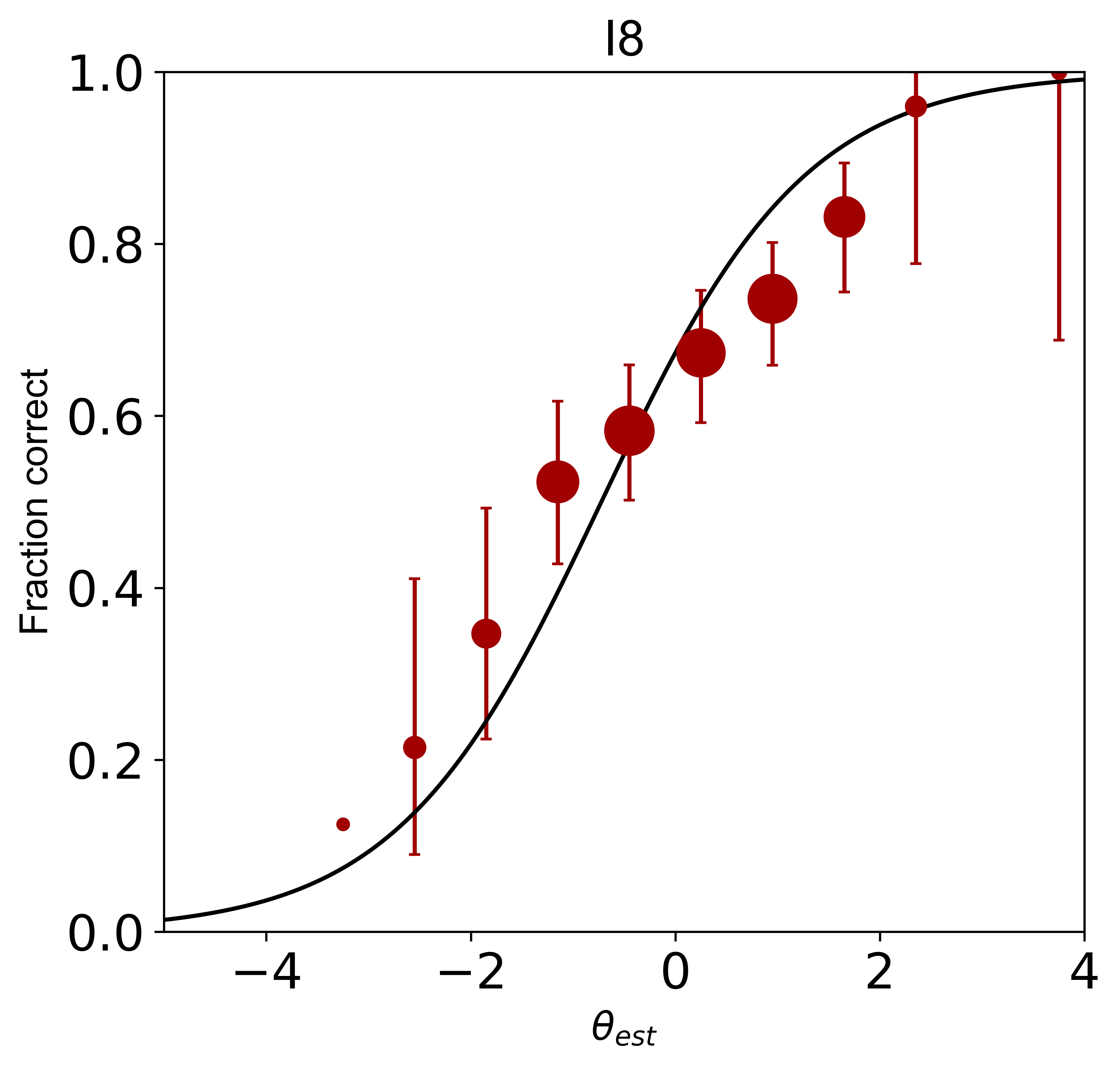}
    \caption{Theoretical item response function (black) and empirical data (red) for Item 8. The size of the marker is proportional to the number of students in that bin. The error bars show a 95\% confidence interval. Error bars are not shown for bins with fewer than 5 students.}
    \label{fig:I8}
\end{figure}

Table~\ref{tab:rasch_fit} shows the infit and outfit statistics for all items. The infit statistics range from 0.88 to 1.18, and the outfit statistics range from 0.70 to 1.29. None of the items have fit statistics falling outside the discussed range. However, Item 8 has fit statistics at the higher end of the range, so it is worth further considering the item. In Fig.~\ref{fig:I8}, we can see the theoretical item response curve and empirical data for Item 8. Consistent with the results of the CTT analysis, the empirical data exhibits a slightly shallower slope than predicted, explaining the higher fit statistics. Nevertheless, the fit statistics for Item 8 are still within the accepted range, and there is no clear evidence that the item is malfunctioning and compromising measurement quality. As such, we retain Item 8, but we will continue to monitor its performance in future administrations. 

\begin{table}[b]
    \centering
    \begin{tabular}{c c c | c c c}
    \hline \hline
    Item & Infit & Outfit & Item & Infit & Outfit \\
    \hline
    \textbf{I1} & 1.01 & 1.01 & \textbf{I11} & 1.00 & 0.99 \\
    \textbf{I2} & 0.97 & 0.96 & \textbf{I12} & 1.00 & 0.97 \\
    \textbf{I3} & 0.94 & 0.84 & \textbf{I13} & 0.92 & 0.82 \\
    \textbf{I4} & 1.04 & 1.06 & \textbf{I14} & 1.06 & 1.10 \\
    \textbf{I5} & 1.01 & 1.03 & \textbf{I15} & 1.03 & 1.05 \\
    \textbf{I6} & 1.01 & 0.99 & \textbf{I16} & 0.94 & 0.88 \\
    \textbf{I7} & 1.06 & 1.03 & \textbf{I17} & 0.92 & 0.87 \\
    \textbf{I8} & 1.18 & 1.29 & \textbf{I18} & 1.04 & 1.07 \\
    \textbf{I9} & 0.92 & 0.85 & \textbf{I19} & 0.97 & 0.93 \\
    \textbf{I10} & 0.88 & 0.70 & \textbf{I20} & 1.07 & 1.13 \\
    \hline
    \end{tabular}
    \vspace{10pt}
    \caption{Infit and outfit statistics.}
    \label{tab:rasch_fit}
\end{table}

\subsubsection{Targeting and reliability}

\begin{figure}
    \centering
    \includegraphics[width=\linewidth]{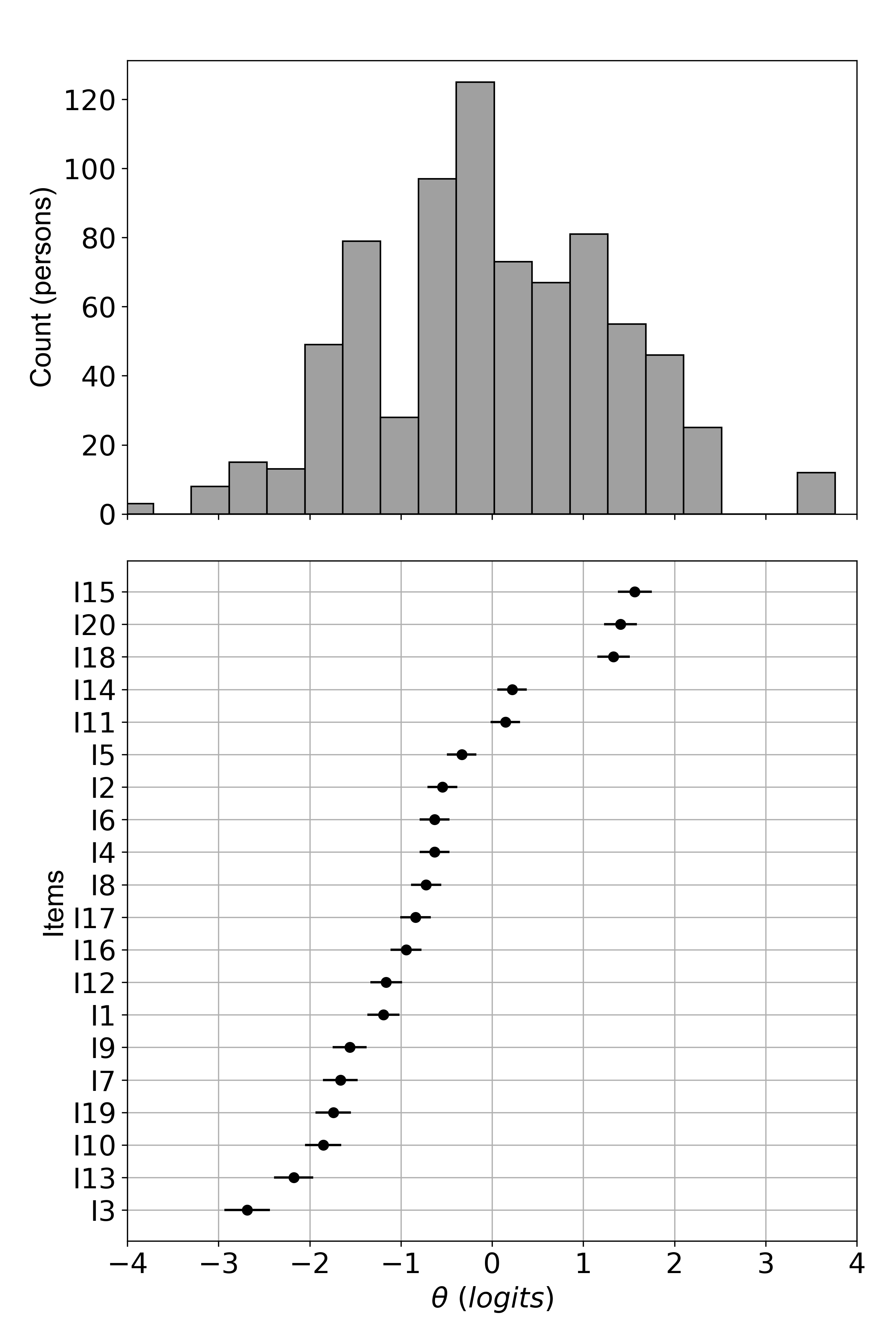}
    \caption{Distribution of student ability estimates (top) and item difficulties (bottom). Item difficulties are shown with 95\% confidence intervals.}
    \label{fig:wright_map}
\end{figure}

In addition to evaluating model fit, it is important to assess whether the instrument appropriately targets the intended population. One visualization to address this question is a Wright map, seen in Fig.~\ref{fig:wright_map}, which shows the distribution of both item difficulties and student abilities on a single plot. Here, it is important to consider whether the range of item difficulties satisfactorily covers the range of student abilities. We note that most of the items have difficulties below 0; since the scale is set such that the mean ability of the sample is 0, this range of item difficulties is consistent with the fact that the instrument was originally designed to target just undergraduate students. Still, we see that the items appropriately span the range of student abilities.  

\begin{figure}
    \centering
    \includegraphics[width=\linewidth]{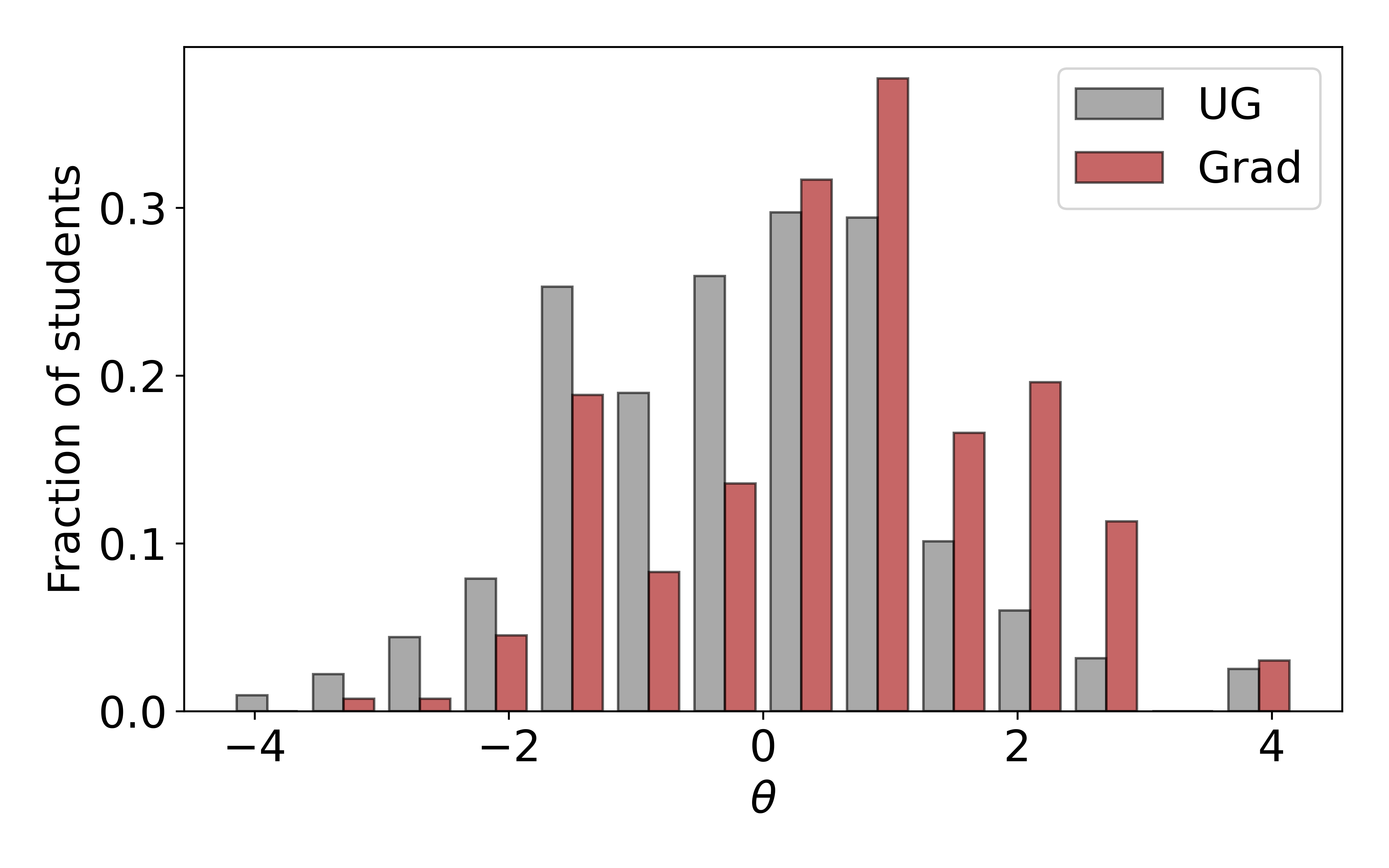}
    \caption{Distribution of $\theta$ for undergraduate and graduate students. Students who did not supply their degree level are not included.}
    \label{fig:grad_ug}
\end{figure}

We can also compare the distribution of undergraduate students to graduate students. While all students are in introductory QIS courses, we would expect the graduate students to score higher, due to a combination of selection effects and previous coursework. Figure~\ref{fig:grad_ug} shows the distribution of $\theta$ for undergraduate and graduate students. We note that the distribution of graduate students is higher, with a mean of 0.49 compared to -0.21, a difference corresponding to a medium effect size ($d = 0.52$, $p < 0.001$). This provides evidence based on the convergence of the QCCS and degree level, a relationship we would expect to observe. 

\begin{figure}
    \centering
    \includegraphics[width=\linewidth]{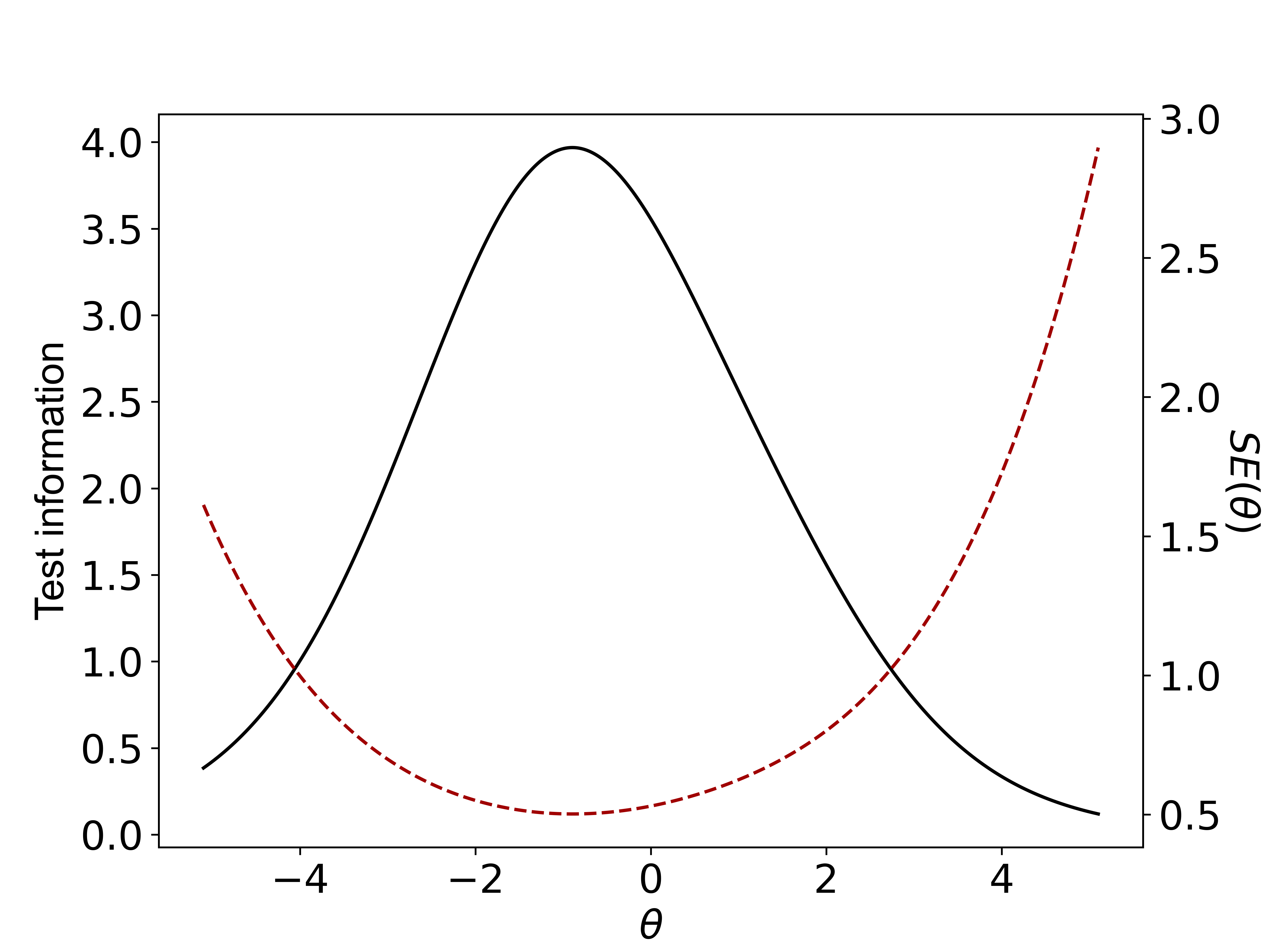}
    \caption{The test information function (black) and standard error (dashed red) as a function of $\theta$.}
    \label{fig:info_se}
\end{figure}

For applications as a measurement tool, we also assess how precisely the instrument measures ability across the latent continuum. For each item, we can calculate an information function, where the information of an item can be intuitively understood as a measure of the information a given item provides toward refining the estimate of latent ability $\theta$. Items have the most information, and thus provide the most precise measurement, at ability levels near their difficulty. By summing each of the individual item information functions, we find the test information function shown in Fig.~\ref{fig:info_se}. Using the test information, we can calculate the standard error of measurement (SEM) as a function of theta. The SEM, shown by the dashed line in Fig.~\ref{fig:info_se}, has a minimum just below $\theta = 0$, and ranges from approximately 0.2 to 1 logit in the range $-4 < \theta <3$, which is consistent with the item distribution seen in Fig.~\ref{fig:wright_map}. With the SEM, we can also calculate a person reliability coefficient analogous to Cronbach's $\alpha$ in CTT. We calculate a reliability of 0.79, which meets the discussed threshold for group measurements and is just below the threshold for individual measurements. As such, it is important to consider if and when conclusions are appropriate for small courses, as are common in QIS.

\subsection{Differential item functioning}\label{sec:dif-detail}

As discussed in Section~\ref{sec:methods}, our preliminary DIF analysis will focus on the axes of race and gender. Without including major and degree level in our model, we theoretically risk confounding effects (e.g., we observe an effect across gender that is actually driven by major composition); however, we consider this to be a reasonable first-order approximation due to the weak relationships between demographic and degree variables, shown in Table~\ref{tab:dem_corr}.

\setlength{\tabcolsep}{5pt}
\begin{table}[b]
\centering
\begin{tabular}{lcccccc}
\hline \hline
 & BS & PhD & Physics & CS & ECE & Math \\
\hline
Man & -0.12 & 0.06 & -0.04 & 0.03 & 0.03 & 0.01 \\
URM & -0.03 & 0.01 & -0.07 & -0.12 & 0.11 & 0.01 \\
\hline
\end{tabular}
\caption{Correlation between demographic groups and majors/degrees. In addition, the correlation between Man and URM is 0.02. Correlations are sufficiently small to suggest that the confounding effects due to degree level or major will be minimal in this analysis.}
\label{tab:dem_corr}
\end{table}

For our analysis, we consider two dichotomized groupings based on gender (man/not-man) and race (URM/not-URM), constructed from self-reported demographic data. We have excluded students who did not respond to the relevant demographic questions. We recognize that this is a limited operationalization of these identities (see e.g. \cite{Traxler:2016}); with increased sample sizes, future work may be able to consider more disaggregated demographic groupings. Based on convention, we treat Men and non-URM students as the reference groups and non-Men and URM students as the focal groups. 

Consistent with our use of the Rasch model, we evaluate DIF using Lord's $\chi^2$ test, an IRT-based procedure. Although we also examined DIF using several alternative methods, we present only the results from Lord's $\chi^2$ test using the 1PL model because the conclusions were unchanged. With this method, item parameters are estimated separately for the reference and focal groups and then linked to a common scale. A $\chi^2$ statistic is then calculated based on the null hypothesis that the linked item parameters are equal between groups. This procedure is used to identify uniform DIF, which is characterized by a constant difference between the reference and focal groups across ability, i.e., the item response functions differ by a constant shift. We calculate p-values from the $\chi^2$ distribution and apply the Benjamini-Hochberg correction to account for multiple testing \cite{Thissen:2002}. The magnitude of DIF is quantified as the difference in estimated item difficulty between the reference and focal groups. We report this difference both in logits and on the ETS scale ($\Delta$), a linear transformation of the logit difference that is widely used in operational DIF analyses because it provides established benchmarks for interpreting effect size \cite{Zwick:2012}. Following the ETS guidelines, $\Delta < 1$ indicates a negligible effect, $1 \leq \Delta < 1.5$ indicates a moderate effect, and $\Delta \geq 1.5$ indicates a large effect \cite{Zwick:2012}.

\begin{figure*}
    \centering
    \includegraphics[width=\linewidth]{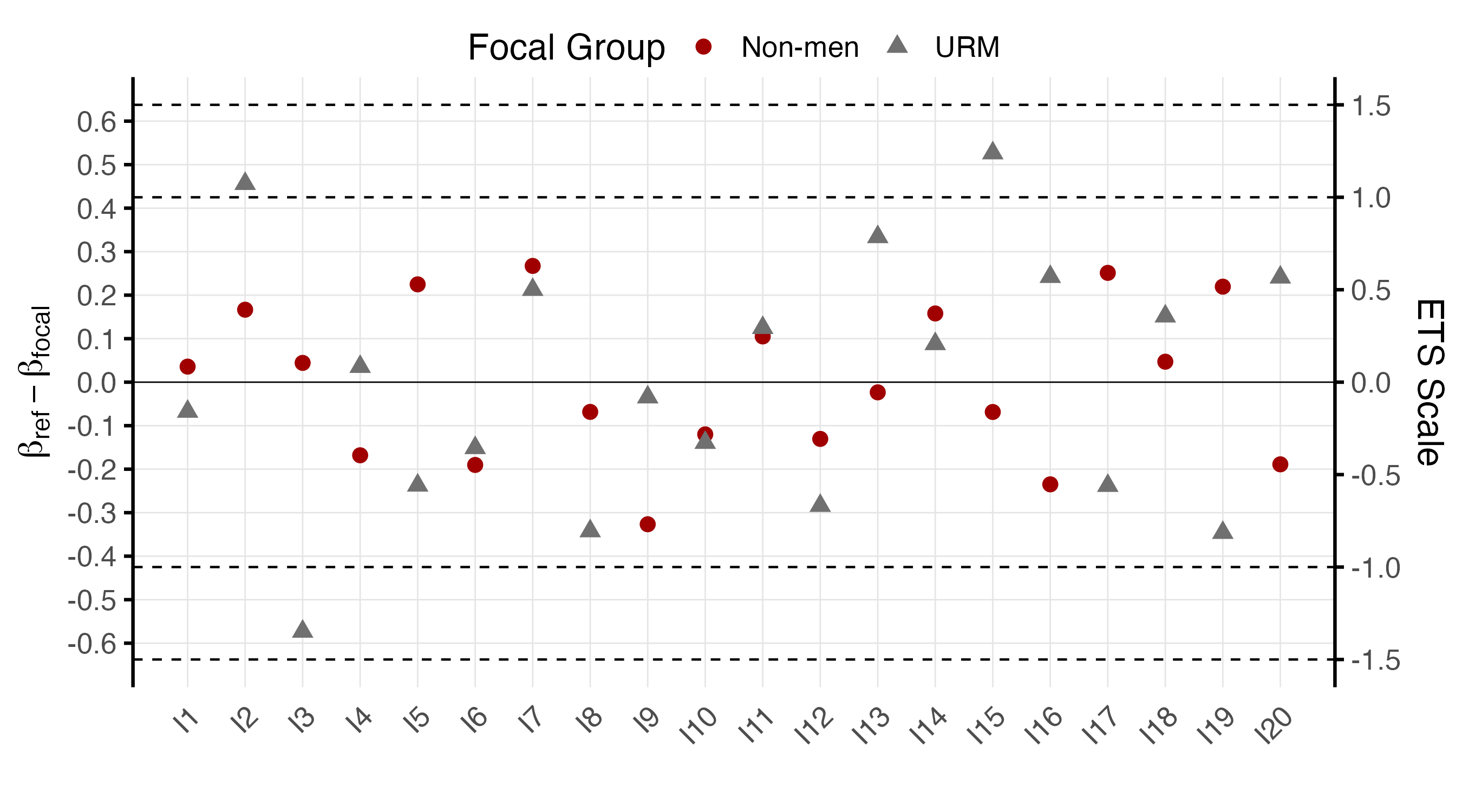}
    \caption{DIF contrasts for gender (non-Men vs. Men) and race (URM vs. non-URM). Points represent the difference in estimated item difficulty between the reference and focal groups, reported in logits (left axis) and on the ETS scale (right axis). Positive values indicate items that are easier for the focal group, whereas negative values indicate items that are harder for the focal group. Dashed horizontal lines indicate the ETS thresholds for moderate ($|\Delta|$ = 1.0) and large ($|\Delta|$ = 1.5) DIF.}
    \label{fig:dif_items}
\end{figure*}

The results of the DIF analysis are shown in Figure~\ref{fig:dif_items}. Along the axis of gender, no items were flagged as exhibiting statistically or practically significant uniform DIF. Along the axis of race, Items 2, 3, and 15 exhibited moderate DIF, but none of these were statistically significant at $\alpha = 0.05$. Furthermore, the observed DIF is not consistently in the same direction—two of the three items exhibit DIF in favor of the focal group. Although these findings warrant continued monitoring, we do not at this stage interpret the results as indicating a substantive threat to fairness or validity.

\section{Discussion}\label{sec:discussion}

In this study, we have discussed the development of the Quantum Computing Conceptual Survey (QCCS) and presented evidence toward its validity and reliability as an instrument for measuring student understanding of quantum computing fundamentals. We have presented both qualitative and quantitative evidence, drawing on Classical Test Theory and Rasch modeling. We will first summarize the evidence presented and outline directions for continued evaluation of the instrument. Then, we will discuss the implications of this work for researchers and instructors. The present paper centers on the validity argument for the QCCS. Unlike many conceptual assessments developed in more mature research areas, the QCCS was developed alongside an emerging body of research on student understanding in quantum computing. As a result, an analysis of how students' response patterns relate to their underlying reasoning—and the resulting implications for instruction—warrants dedicated treatment, which we pursue in a separate, forthcoming paper.

\subsection{Evidence for validity and reliability}

Here, we reiterate the evidence for validity and reliability presented throughout the paper, as summarized in Table~\ref{tab:evidence}. 

\emph{Evidence based on test content:} The topics and assessment objectives were informed by extensive input from QIS instructors, domain experts, and QIS education researchers to ensure the QCCS adequately represents the intended construct domain. Individual items were developed based on prior research into student difficulties with quantum computing fundamentals and closely related topics. 

Given the rapidly evolving nature of QIS education, ongoing evaluation of the instrument's content will be essential. Future work will involve considering whether the QCCS continues to emphasize topics considered important by instructors and the broader field, as well as exploring possible additional topics to better target more advanced classes. 

\emph{Evidence based on response process:} Item development and revision were guided by think-aloud interviews with QIS students from a variety of disciplines and degree programs. In addition, many item distractors were derived from student responses to an earlier free-response version of the instrument, providing further insight into students' reasoning.  

With the existing qualitative data, we have been conducting more detailed analyses of student cognitive processes and intend to continue this work \cite{Meyer:2026, Plueger:2026}. We also envision the QCCS as a tool for investigating learning progressions within QIS (e.g., see \cite{Neumann:2013}) and for examining how these progressions may vary across students with different academic backgrounds. 

\emph{Evidence based on internal structure:} We presented evidence supporting the unidimensionality of the instrument, justifying the use of a single score. We also demonstrated sufficient reliability for group-level comparisons. All items exhibited positive and sufficiently large discriminations, and we do not observe any items with concerning misfit to the Rasch model. We also examined differential item functioning across the axes of race and gender and did not observe evidence of concerning, uniform DIF across these groups. 

Future work will include a more detailed study of potential DIF, particularly with respect to students in different majors. Until such analyses are conducted, comparisons of scores across these student groups should be interpreted with caution. 

\emph{Evidence based on relations to other variables:} We examined differences in ability between undergraduate and graduate students and found results that align with theoretical expectations. 

With other instruments, a common source of supporting validity evidence is the correlation between the instrument score and an external, related measure (e.g., an existing RBA). At present, there is no well-established QIS metric against which to conduct such an analysis. This remains an open consideration and should be revisited as appropriate comparison measures become available. 

\subsection{Implications for researchers}

\subsubsection{Implications for QIS education research}

Given the relative newness of the field, many open questions remain to be explored within QIS education research. Beyond the immediate benefits to instructors, our team prioritized the early development of the QCCS in part as an accelerator for QIS education research. We hope that the rich quantitative datasets produced by the QCCS will serve as a productive resource education researchers can utilize for QIS education research moving forward.

Analyses of student responses can reveal common student difficulties and naive conceptions about quantum computing fundamentals, providing a foundation for targeted qualitative studies. Examining how item-level performance and distractor selection vary across student backgrounds, prior coursework, and course contexts can illuminate the influence of interdisciplinary backgrounds and environments on learning outcomes. Comparison with existing quantum mechanics RBAs could further clarify the relationship between conceptual proficiency in quantum mechanics and quantum computing. These insights may inform the design of instructional interventions, curricular materials, or course structures tailored to diverse student populations. Additionally, interview studies with instructors in particularly high- and low-performing courses can shed light on pedagogical strategies that support learning in QIS contexts. 

The QCCS also provides a natural benchmark for empirically evaluating the effectiveness of curricular instruments (e.g., tutorials, clicker question sequences). Similarly, reliable metrics for measuring QCCS fundamentals open the door to quasi-controlled studies on active research questions in QIS pedagogy, such as the relative merits of various two-state systems as qubit archetypes for teaching \cite{Bitzenbauer:2024}, advantages and drawbacks of different quantum state representations \cite{Rexigel:2026}, or the effectiveness of Quantum in Pictures (ZX calculus) vs.\ the traditional quantum circuit model in introducing quantum computing to new audiences \cite{Dundar-Coeke:2023}. More broadly, the QCCS serves as a reliable measurement tool to support progress in QIS education research and pedagogy, much as RBAs have historically done for PER in other subjects. 

\subsection{Implications for instructors}

We encourage faculty teaching quantum computing and quantum information coursework to use the QCCS as a tool to measure the effectiveness of teaching and curricula, provided the content and learning goals of the course appropriately match those of the assessment. Of course, we caution against over-reliance on any single metric and recommend considering QCCS scores as part of a broader holistic evaluation of teaching effectiveness.

Instructors may wish to use the QCCS for the following:
\begin{itemize}\setlength\itemsep{0.5em}
    \item Comparing student performance in your course to similar courses taught by different instructors and/or at different institutions
    \item Evaluating student understanding of fundamentals before moving on to more advanced topics
    \item Considering the effectiveness of changes to curricula via pre-/post-studies
    \item Tracking student performance in your courses over time
    \item Identifying and remedying problematic conceptions that students may still be holding after instruction
\end{itemize}

Because QIS and QIS education are both evolving rapidly, course content, learning goals, and pedagogical approaches are still very much in flux across instructors and institutions. As such, instructors are required to make decisions about what content to include, emphasize, or omit, sometimes with limited prior evidence to guide those choices. The QCCS can help support these decisions and help instructors evaluate whether their instructional choices are reaching their intended goals. In addition, the QCCS can serve as evidence when communicating with departments, institutions, or other stakeholders, enabling instructors to advocate for the value, rigor, and continued development of QIS courses and programs. 

\section{Acknowledgments}

Special thanks to our student interviewees and survey respondents, and to the faculty who contributed to the development of QCCS through scope analysis, preliminary validation interviews, and pilot testing. We would like to acknowledge Giaco Corsiglia and Bianca Cervantes for participation in early exploratory studies, as well as Jonan-Rohi Plueger and Michael Burnes for assistance with item refinement. This work is supported by the University of Colorado Boulder Department of Physics, the NSF Graduate Research Fellowship Program, and NSF Grants Nos.\ 2011958, 2012147, and 2143976.
\FloatBarrier

\appendix

\section{Assessment objectives}\label{sec:appAO}

\begin{table*}[tb]
    \centering
    \begin{tabular}{l p{350px} l} \hline\hline
    \thead{AO} & \thead{Students should be able to ...} & \thead{Associated items}
   \\ \hline \\
    1 & Convert between column vector and bra-ket notation & 2\\
    2 & Compute the inner product of two quantum states; identify its dimension as a scalar (or meaningless if different Hilbert space) & 4, 20\\
    3 & Compute the tensor product of two 1-qubit quantum states in column vector or bra-ket notation; identify its dimension & 4, 7, 9, 19\\
    4 & Determine whether an expression results in a scalar, bra, ket, operator, or is meaningless & 4, 9\\
    5 & Determine the dimension of the Hilbert space of a system of N qubits & 2\\
    \textit{6} & \textit{Identify whether a matrix is unitary} & -- \\
    7 & Determine (by eye/naive measurement probabilities \cite{Meyer:2021PERC}) whether a simple 2-particle quantum state is entangled or non-entangled. If non-entangled, write as tensor product of 1-particle states & 7, 12\\
    8 & Given a multi-qubit state, find the state of a subset of qubits (or state that it is not well-defined as a ket) & 15, 18\\
    \textit{9} & \textit{Qualitatively identify entanglement (and the exponential increase in the size of the Hilbert space) with quantum advantage} & -- \\
    10 & Compute measurement probabilities on a 1 or 2-particle superposition state & 10 \\
    11 & Write the state of a 1- or 2-particle superposition state before and after measurement (including partial measurement), or state that it is ill-defined & 15\\
    \textit{12} & \textit{Qualitatively distinguish between a superposition state and classical randomness} & -- \\
    13 & Evaluate the effects of standard gates* (I, X, Z, H, CNOT) on 1- and 2-qubit quantum states & 1, 6, 8, 14, 15\\
    14 & Determine whether two gates (on the same or different qubits) commute & 1 \\
    15 & Qualitatively determine whether a 2-qubit gate produces entanglement between two previously unentangled qubits & 16 \\
    16 & Interpret the tensor product of two single-qubit quantum gates & 1, 3, 13, 16 \\
    \textit{17} & \textit{Identify that a ``black box'' quantum gate must be reversible/unitary to be physically realizable} & -- \\
    18 & Compute the final state of a 1- or 2-qubit quantum circuit given a starting state and sequence of gates & 6, 15\\
    19 & Convert between matrix, algebraic, and quantum circuit diagram notations for single qubit gates and for 2 qubits &  11, 17 \\
    20 & Identify whether two quantum circuit diagrams are equivalent; simplify quantum circuit diagrams using elementary algebraic properties of 1-qubit gates and CNOT & 1, 8 \\
    \hline\hline
    \end{tabular}
    \caption{Assessment objectives formulated for QCCS. Italicized AOs were dropped in later stages of item development and refinement, motivated either by feedback from instructors or by iterative removal of all corresponding items due to low statistical performance. *Y gate originally included in this list but dropped from final assessment due to conceptual redundancy with X and Z.}
    \label{tab:initial-AOs}
\end{table*}

\section{Pilot population characteristics}\label{sec:appA}

\begin{table*}[h]
    \centering
    \begin{tabular}{l c c c}
        \hline \hline
         \thead{} & \thead{v2.0} & \thead{v2.1} & \thead{v2.2}\\
         \hline
         Usable student responses        & 621       & 345 & 777 \\
         \multicolumn{3}{l}{\ \ \textit{Student self-reported race/ethnicity}} \\
         \textit{Reported}  &                \textit{578 (93\%)} & \textit{319 (92\%)}  & \textit{705 (91\%)}  \\
         Asian                              & 53\%      & 36\%   & 44\%                \\
         Black/African American             &  2\%      &  3\%  & 4\%            \\
         Caucasian                          & 39\%      & 50\%  & 46\%          \\
         Hispanic/Latinx                    &  9\%      & 16\%   & 12\%            \\
         Middle Eastern or North African    &  3\%      &  2\%   & 5\%            \\
         Native American                    &<1\%       & <1\%   & <1\%            \\
         Pacific Islander                   &<1\%       & <1\%  & <1\%          \\
         Any URM\footnote{\textit{URM} includes any of the following (self-report): Black/African American, Hispanic/Latinx, Middle Eastern or North African, Native American, or Pacific Islander.}                            & 15\%      & 20\%  & 21\%           \\
         \multicolumn{3}{l}{\ \ \textit{Student self-reported gender}} \\
         \textit{Reported}  &              \textit{590 (95\%)} & \textit{326 (94\%)}  & \textit{716 (92\%)} \\
         Man                                & 79\%      & 82\%  & 78\%                  \\
         Woman                              & 19\%      & 17\%  & 20\%               \\
         Non-binary                         &  3\%      &  1\%   &2\%               \\
        \multicolumn{3}{l}{\ \ \textit{Student self-identifies as disabled and/or neurodivergent}} \\
        \textit{Reported}  &              \textit{559 (90\%)} & \textit{303 (88\%)} & \textit{680 (88\%)}  \\
        Yes                                 &  9\%       & 16\%   & 16\%               \\
        No                                  & 91\%       & 84\%  & 84\%                  \\
        \multicolumn{3}{l}{\ \ \textit{Student self-identifies as first generation and/or from low income family}} \\
        \textit{Reported}  &              - & \textit{313 (91\%)} & \textit{705 (91\%)} \\
        Yes                                 & -          & 28\% & 28\% \\
        No                                  & -          & 72\% & 72\%\\
        \multicolumn{3}{l}{\ \ \textit{Student self-reported highest degree currently obtaining}} \\
        \textit{Reported}  & \textit{614 (99\%)}         & \textit{340 (99\%)}  & \textit{753 (97\%)} \\
        Bachelor's                          & 58\%       & 55\%   & 70\% \\
        Master's                            & 27\%       & 16\%   & 15\% \\
        Ph.D.                               & 14\%       & 28\%   & 14\%\\
        Other                               & <1\%       & <1\%   & <1\%\\
        \multicolumn{3}{l}{\ \ \textit{Student self-reported major}} \\
        \textit{Reported}  & \textit{615 (99\%)}        & \textit{340 (99\%)} & \textit{752 (97\%)} \\
        Physics                             & 28\%      & 49\%     & 34\% \\
        CS                                  & 40\%      & 31\%   & 39\% \\
        ECE                                 & 32\%      & 21\%  & 24\%  \\
        Math                                & 18\%      & 10\%    & 16\% \\
        Other\footnote{\textit{Other} only includes students who did not report one of Physics, CS, ECE, or Math.}                               &  6\%      & 6\%    & 6\% \\
        \hline \hline
         
    \end{tabular}
    \caption[Demographic profiles for students in the v2.0, v2.1, and v2.2 pilot pools.]{Demographic profiles for students in the v2.0, v2.1, and v2.2 pilot pools. Percentages are of students who opted to report the given identity, and may not add up to 100\% due to rounding and/or non-exclusive categories.}
    \label{tab:student-demographics}
\end{table*}

\FloatBarrier

\section{CTT Statistics}\label{sec:appB}

\begin{table}[ht!]
    \centering
    \begin{tabular}{c | c c }
    \hline \hline
    \thead{Item} & \thead{Difficulty} & \thead{Discrimination} \\
    \hline
    I1 & 0.72(2) & 0.41(3) \\
I2 & 0.60(2) & 0.47(3) \\
I3 & 0.90(1) & 0.37(3) \\
I4 & 0.62(2) & 0.40(3) \\
I5 & 0.56(2) & 0.42(3) \\
I6 & 0.62(2) & 0.42(3) \\
I7 & 0.79(1) & 0.35(3) \\
I8 & 0.64(2) & 0.26(3) \\
I9 & 0.77(1) & 0.48(3) \\
I10 & 0.81(1) & 0.51(3) \\
I11 & 0.47(2) & 0.43(3) \\
I12 & 0.71(2) & 0.42(3) \\
I13 & 0.85(1) & 0.44(3) \\
I14 & 0.46(2) & 0.37(3) \\
I15 & 0.23(2) & 0.32(3) \\
I16 & 0.68(2) & 0.49(3) \\
I17 & 0.66(2) & 0.50(3) \\
I18 & 0.26(2) & 0.33(3) \\
I19 & 0.80(1) & 0.41(3) \\
I20 & 0.25(2) & 0.30(3) \\
    \hline \hline
    \end{tabular}
    \caption{CTT item difficulties and discriminations for v2.2. Discrimination is calculated as the point-biserial correlation. Parenthetical uncertainties denote standard error, calculated with 1000 bootstrapped samples, and are with respect to the least significant digit.}
    \label{tab:ctt-diff-disc}
\end{table}

\FloatBarrier

\bibliography{onebib}

\end{document}